\documentclass[journal=ancac3,manuscript=letter,etalmode=truncate,maxauthors=100]{achemso}

\usepackage[version=3]{mhchem} 
\usepackage{textcomp}
\usepackage{xcolor}
\usepackage{soul}
\usepackage{dirtytalk}
\usepackage{microtype}
\usepackage{verbatim}
\usepackage{siunitx}
\usepackage{wrapfig}
\usepackage{multirow}
\usepackage{array}
\usepackage{gensymb}
\usepackage[utf8]{inputenc}
\usepackage{amssymb}
\usepackage{pdflscape}
\usepackage{booktabs}
\usepackage{graphicx}

\usepackage{pdfpages}

\newcommand{\supplementarysection}{%
  \setcounter{figure}{0}
  \let\oldthefigure\thefigure
  \renewcommand{\thefigure}{S\oldthefigure}
  \setcounter{table}{0}
  \let\oldthetable\thetable
  \renewcommand{\thetable}{S\oldthetable}
  \section{Supplementary section}
}

\setkeys{acs}{etalmode=truncate,maxauthors=10}

\SectionNumbersOn

\author{Martin Jindra}
\affiliation[JHI]
{Heyrovský Institute, Czech Academy of Sciences, Dolejškova 2155/3, 182 00 Prague 8, Czech Republic}
\alsoaffiliation[UCT]
{Department of Physical Chemistry, University of Chemistry and Technology in Prague, Technická 3, 166 28 Prague 6, Czech Republic}

\author{Luka Pirker}
\affiliation[JHI]
{Heyrovský Institute, Czech Academy of Sciences, Dolejškova 2155/3, 182 00 Prague 8, Czech Republic}

\author{Tim Verhagen}
\affiliation[FZU]
{Institute of Physics, Czech Academy of Sciences, Na Slovance 1999/2, 182 00 Prague 8, Czech Republic}

\author{Ciarán Fowley}
\affiliation[HZDR]
{Institute of Ion Beam Physics and Materials Research, Helmholtz-Zentrum Dresden - Rossendorf, Bautzner Landstrasse 400, 01328 Dresden, Germany}

\author{Martin Kalbáč}
\affiliation[JHI]
{Heyrovský Institute, Czech Academy of Sciences, Dolejškova 2155/3, 182 00 Prague 8, Czech Republic}

\author{Matěj Velický}
\affiliation[JHI]
{Heyrovský Institute, Czech Academy of Sciences, Dolejškova 2155/3, 182 00 Prague 8, Czech Republic}

\author{Riichiro Saito}
\affiliation[NTNU]
{Department of Physics, National Taiwan Normal University, Heping E. Rd., Taipei 106, Taiwan}
\alsoaffiliation[TU]
{Department of Physics, Tohoku University, Sendai, 980-8578, Japan}

\author{Ado Jorio}
\affiliation[UFMG]
{Departamento de F\'isica, Universidade Federal de Minas Gerais, Belo Horizonte, Minas Gerais 30123-970, Brazil}

\author{Otakar Frank}
\affiliation[JHI]
{Heyrovský Institute, Czech Academy of Sciences, Dolejškova 2155/3, 182 00 Prague 8, Czech Republic}
\email{otakar.frank@heu.cas.cz}

\title[SaS]
  {Beyond Equilibrium Raman Thermometry: Fermi-Level Control of the Anti-Stokes Response in Graphene}


\begin{document}


\begin{tocentry}

\includegraphics{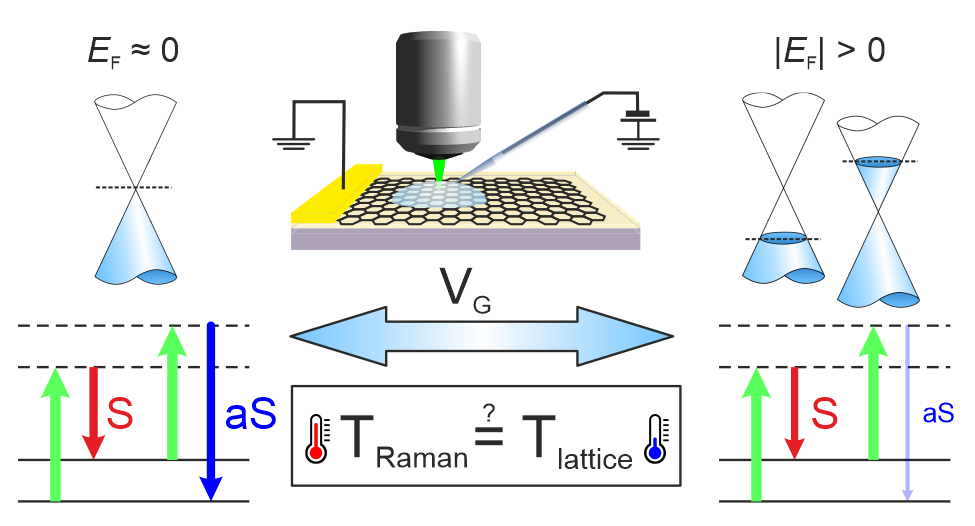}
 
\end{tocentry}

\begin{abstract}

Non-equilibrium electron and phonon populations govern heat generation and dissipation in nanoscale devices, yet their direct characterization remains challenging. Anti-Stokes Raman scattering provides a sensitive probe of phonon populations and local temperatures, although its interpretation becomes more complex when electronic and phononic populations are driven out of equilibrium. Here, we investigate the gate-dependent anti-Stokes Raman response of monolayer graphene and find that the anti-Stokes response deviates from that expected from an equilibrium phonon population. Using a spatially localized microdroplet spectroelectrochemical platform, dielectric back-gates, and chemical doping, we observe a strong and reversible suppression of the G‑mode anti‑Stokes intensity as the Fermi level is tuned away from the Dirac point, while the Stokes intensity remains essentially unchanged. By contrast, the G-mode frequency and linewidth of both the Stokes and anti-Stokes branches evolve with doping as expected from the established Kohn-anomaly picture. The anti‑Stokes/Stokes intensity ratio decreases by several-fold over the measured range and displays a pronounced, gate‑dependent power response with maximum sensitivity at charge neutrality. Comparison with previous studies of non-equilibrium carrier dynamics in doped graphene suggests an important role of Fermi-level-dependent electronic phase space, which can influence both hot-carrier-driven phonon populations and electronically mediated Stokes–anti-Stokes scattering. Our findings establish anti‑Stokes Raman scattering as a sensitive probe of electron–phonon coupling and non‑equilibrium phonon dynamics in graphene.


\end{abstract}


\newpage

\section*{Introduction}

Controlling and understanding energy dissipation in low-dimensional materials is central to the performance of nanoscale electronic, optoelectronic, and quantum devices, where efficient heat removal from confined electron and phonon subsystems can be the limiting factor for device speed, stability, and functionality.\cite{balandin2011,pop2012} In two-dimensional (2D) materials and van der Waals heterostructures, this challenge is especially acute: reduced dimensionality concentrates absorbed energy into a small number of atomic layers, and non-equilibrium electron and phonon populations can persist far longer than in bulk systems before relaxing back to the lattice temperature.\cite{breusing2011} Recent work has shown that cooling pathways in 2D systems can be highly tunable and even exploited constructively, from doping-dependent hot-carrier relaxation in graphene \cite{graham2013} to interfacial-charge-transfer-driven optical refrigeration in semiconductor heterobilayers.\cite{lin2026}

Monolayer graphene, with its linear electronic dispersion and continuously tunable Fermi level ($E_\textrm{F}$),~\cite{novoselov2004, neto2009} provides a particularly well-defined platform for dissecting the non-equilibrium electron–phonon dynamics. This electron-phonon coupling is especially pronounced for the G phonon mode, whose frequency and linewidth are strongly renormalized by the associated Kohn anomaly -- a consequence of the mode's coupling to electron–hole excitations.~\cite{lazzeri2006} As a result, Raman spectroscopy of the G mode~\cite{ferrari2006, malard2009} has been established as a sensitive, non-invasive probe of doping and electron–phonon coupling strength in graphene.~\cite{lazzeri2006, piscanec2004, neto2007}

Raman thermometry attributes the anti-Stokes-to-Stokes integrated intensity ratio ($I_{\mathrm{aS}}/I_{\mathrm{S}}$) to the thermal phonon population and is therefore commonly used to extract an effective phonon temperature.~\cite{faugeras2010,tristant2016} However, a growing body of work has shown that the electronic and optical-phonon subsystems in graphene under optical or electrical excitation can depart substantially from the lattice equilibrium, which limits a purely thermal interpretation of the anti-Stokes response.~\cite{berciaud2010,graham2013,Ferrante2018} This caveat is not unique to graphene: recent work on van der Waals heterostructures has shown that anomalous $I_{\mathrm{aS}}/I_{\mathrm{S}}$ signatures can also arise from genuinely non-thermal, charge-transfer-driven processes, and has cautioned that resonance effects can further complicate a straightforward thermometric interpretation of the anti-Stokes response.\cite{lin2026}

Several complementary methods provided evidence for this non-equilibrium behavior. Ultrafast optical spectroscopy revealed rapid carrier thermalization followed by cooling through optical-phonon emission and hot-phonon effects,~\cite{breusing2011} while time- and angle-resolved photoemission showed that the transient electronic temperature and cooling dynamics depend strongly on doping.~\cite{johannsen2015,weigl2026} Terahertz spectroscopy and photocurrent studies reached similar conclusions, showing that hot-electron temperatures are highest near charge neutrality and that the cooling dynamics depend on both the electronic heat capacity and electron–phonon scattering.~\cite{lin2015,graham2013} Consistently, Bistritzer and MacDonald emphasized the role of graphene’s Dirac density of states in electronic cooling,~\cite{bistritzer2009} and electrical biasing of graphene showed that the electronic and G-mode optical-phonon temperatures closely track each other while lower-energy phonons may remain out of equilibrium.~\cite{berciaud2010}

These results imply that the anti-Stokes signal in graphene need not be governed solely by an equilibrium phonon population. At the same time, the electron–phonon coupling relevant to the G mode is itself strongly dependent on $E_\textrm{F}$: it is maximal near charge neutrality (Dirac point) and suppressed away from it through Pauli blocking, as established previously in Raman studies.~\cite{yan2007,das2008,ando2006,hasdeo2014,lazzeri2006} More recently, spatially coherent tip-enhanced Raman measurements demonstrated a pronounced Fermi-level dependence of the Raman coherence length in gated graphene, providing further evidence that the carrier concentration strongly modifies the spatial and dynamical characteristics of the Raman scattering process.~\cite{nadas2023} Together with earlier observations of enhanced Stokes--anti-Stokes correlation under strong optical excitation,~\cite{kasperczyk2015,jorio2015} this suggests that the electronic continuum may actively mediate the anti-Stokes process in graphene.

Accessing this regime experimentally requires simultaneously achieving a wide, continuously tunable doping range and sufficient optical sensitivity to detect the intrinsically weak G-mode anti-Stokes signal. Dielectric back-gating alone cannot provide the carrier densities needed to cover the needed doping range, while chemical doping does not provide the continuous, in situ reversibility required to track the evolution of $I_{\mathrm{aS}}/I_{\mathrm{S}}$. A thick electrolyte layer, although capable of reaching the needed doping levels, introduces scattering and absorption losses that reduce the already weak anti-Stokes signal to the noise floor. To overcome this trade-off between tuneable doping range and optical sensitivity, we employed a spatially confined microdroplet approach, in which lithographically defined reservoirs confine the electrolyte to a thin layer directly above the probed graphene region. This setup delivers the full electrolyte gating range while minimizing the optical path through the liquid.

We investigate the role of the electronic system in the anti-Stokes response of graphene by examining the G-mode Raman response under controlled Fermi-level tuning using electrolyte, dielectric, and chemical gating. The measurements reveal a strong and reversible suppression of the anti-Stokes intensity away from charge neutrality, together with a pronounced gate dependence of the laser-power-dependent evolution of $I_{\mathrm{aS}}/I_{\mathrm{S}}$. These results indicate that the anti-Stokes response cannot be understood solely in terms of an equilibrium phonon population and instead point to an additional electronic contribution to the anti-Stokes response that depends on the tunable charge-carrier density.


\section*{Results}\label{result}
\subsection*{Localized electrolyte gating}

\begin{figure}[h!]
    \centering
    \includegraphics[width=1\linewidth]{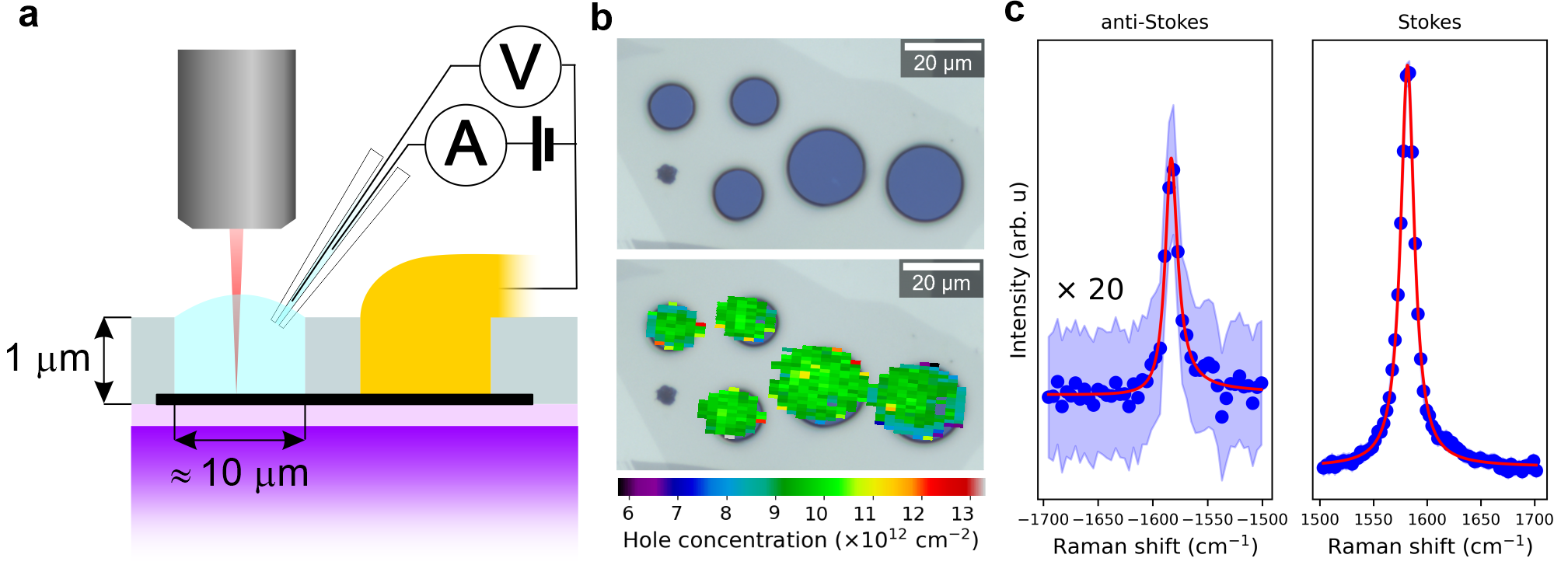}
    \caption{\textbf{Experimental platform for gate-controlled Raman measurements in graphene.} \textbf{a,} Schematic of the localized electrolyte gating configuration with a lithographically defined reservoir for the electrolyte. \textbf{b,} Optical image of a graphene device with micro-patterned reservoirs (top). The charge-carrier density distribution map obtained from the Raman G-2D frequency correlation plot (bottom). \textbf{c,} Comparison of the anti-Stokes (left) and Stokes (right) spectra of the graphene G mode acquired at an applied potential of $-0.4$~V, i.e., at the Dirac point (laser excitation wavelength of 633~nm, 1.5~mW power at the sample, 3.8~M LiClO\textsubscript{4} aq. electrolyte). Blue dots are the experimentally measured points, the red lines show the fitted Breit-Wigner-Fano lineshapes, and the shaded areas represent the signal uncertainty.}
    \label{fig1}
\end{figure}

To investigate the Stokes and anti-Stokes Raman response of graphene under controlled Fermi-level modulation, we employed a spatially localized electrolyte gating scheme. This approach enables reversible control of the carrier density while preserving optical access to the probed region. A schematic of the device geometry is shown in Fig. \ref{fig1}a, where lithographically defined reservoirs confine the electrolyte to a selected area on mechanically exfoliated monolayer graphene (detailed description in Supporting Information Section \ref{Sample_prep-SI}). The optical image in Fig. \ref{fig1}b identifies the region selected for Raman mapping (detailed description in Supporting Information Section \ref{Measurement-SI}). The carrier-density distribution was obtained using the correlation between the G- and 2D-mode frequencies, and the corresponding data cluster close to the iso-doping line of $(9.5\pm 0.7)\times10^{12}$ cm$^{-2}$, with an average tensile strain of $(0.18\pm 0.05) \%$ (Supporting Fig. \ref{map_corr}).
Fig. \ref{fig1}c compares the anti-Stokes and Stokes Raman spectra of the graphene G mode acquired at $-$0.4~V (i.e., at the Dirac point; see below), showing that the anti-Stokes signal is clearly detectable, although substantially weaker than the Stokes counterpart. The intensity of the anti-Stokes 2D was insufficient for analysis.

\subsection*{Gate-dependent Stokes and anti-Stokes response}

\begin{figure}[h!]
    \centering
    \includegraphics[width=1\linewidth]{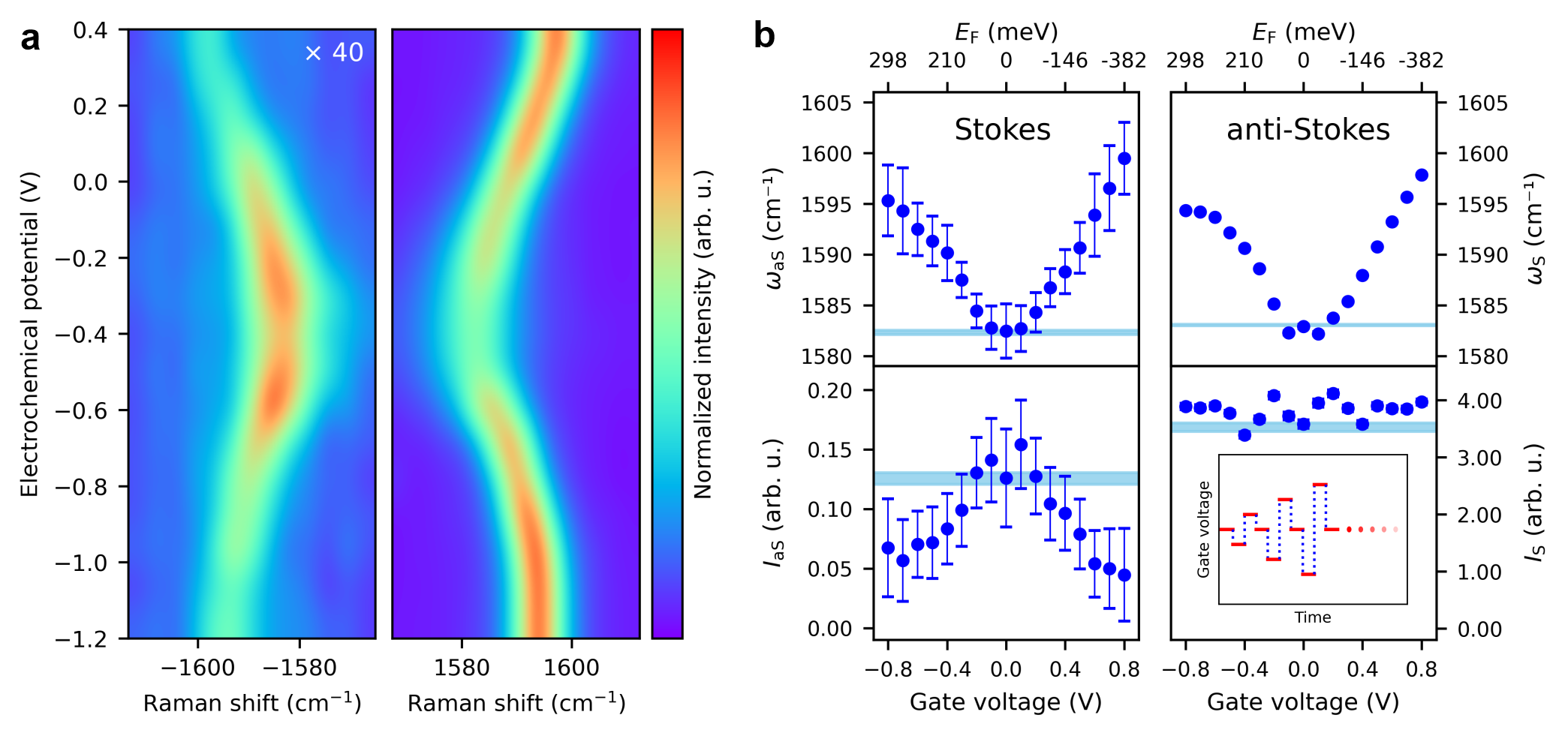}
    \caption{\textbf{Gate-dependent evolution of the anti-Stokes and Stokes G modes of graphene.} \textbf{a,} Intensity maps of the anti-Stokes (left) and Stokes (right) G bands as a function of applied electrochemical potential, which are symmetric with respect to the Rayleigh line (laser excitation wavelength of 633~nm, 1.5~mW power at the sample, 3.8~M LiClO\textsubscript{4} aq. electrolyte). \textbf{b,} Peak positions and peak integrated areas of the anti-Stokes (left panels) and Stokes (right panels) G band. The blue-shaded bands indicate the confidence interval ($\pm$1 standard deviation around the mean) of repeated measurements performed at the Dirac point, demonstrating the reversibility of the gating process. The error bars represent the fitting error, including the weighting in the photon counting uncertainty. The inset in the $I_\textrm{S}$ panel shows the applied gate-voltage protocol as a function of time; Raman spectra were acquired within the red line segments.}
    \label{fig2}
\end{figure}

The gate-dependent evolution of the graphene anti-Stokes and Stokes Raman G bands is summarized in Fig. \ref{fig2}a as Raman intensity heatmaps versus the applied electrochemical potential ($V$). Measurements within the available potential window show that the Stokes frequency reaches a minimum at about $-$0.4~V, identifying the Dirac point, i.e. charge-neutrality point, in this device ($V_\textrm{Dirac}$). At all other potentials, both branches shift to higher absolute Raman frequency. The 2D--G correlation plot, shown in Supporting Fig. \ref{corr_single} for the same experiment, also documents the expected potential-dependent evolution of the 2D band. For consistency with the conventional physics notation, we further define the gate voltage as $V_\textrm{G} = -(V-V_\textrm{Dirac})$. The spectral evolution was quantified by fitting the G band using a Breit-Wigner-Fano line shape (see Supporting Information Section \ref{SI_peakfitting}),~\cite{hasdeo2014} and the extracted anti-Stokes and Stokes peak positions ($\omega_\textrm{aS}, \omega_\textrm{S}$, top panels) and areal intensities ($I_\textrm{aS}, I_\textrm{S}$, integrated under the fitted curve, bottom panels) are plotted in Fig.~\ref{fig2}b as a function of $V_\textrm{G}$ (bottom axes) and the estimated $E_\textrm{F}$ (top axes). The latter was inferred from the G-band shift using the linear approximation given in Eq.~\ref{eq0}.~\cite{chen2011}

\begin{equation}
    \Delta\omega_\textrm{G} = 42 \times |E_\textrm{F}|\ \textrm{cm}^{-1}\textrm{eV}^{-1}
\label{eq0}
\end{equation}

The anti-Stokes and Stokes peak positions (Fig.~\ref{fig2}b, top) both display the expected V-shaped dependence on $V_\textrm{G}$, consistent with the renormalization of the G mode with carrier density.~\cite{das2008} In contrast, their integrated areal intensities behave very differently (Fig.~\ref{fig2}b, bottom): the Stokes intensity remains approximately constant across the gating range, whereas the anti-Stokes intensity reaches a maximum near the Dirac point and decreases as $E_\textrm{F}$ is shifted away from it. The effect is fully reversible, with the G-band parameters returning to their initial values whenever the applied potential is restored to the previously determined Dirac-point potential, as evidenced by the narrow confidence interval of the repeated measurements in Fig.~\ref{fig2}b (blue-shaded bands). The applied gate-voltage protocol is shown in the inset of Fig.~\ref{fig2}b (bottom-right). For reference, the full evolution of amplitudes, calculated areas, linewidths, and asymmetry factors is provided in Supporting Fig. \ref{all_G}. 

\subsection*{Cross-platform consistency}

To test whether the decrease in anti-Stokes G-band areal intensity with doping is intrinsic to graphene rather than specific to the microdroplet configuration, complementary measurements were carried out using different electrolytes, dielectric back-gating (Supporting Fig. \ref{fig:backgate}), and chemical doping (Supporting Fig. \ref{fig:chemgate}). In all cases, the same qualitative trend was observed: the anti-Stokes response decreased with increasing carrier density, whereas the Stokes response remained much less affected. When plotted as $I_{\mathrm{aS}}/I_{\mathrm{S}}$ versus $|E_\mathrm{F}|$ calculated from the G-mode shift as noted above, the data from all gating approaches collapse onto a common $|E_\mathrm{F}|$-dependent trend (Supporting Fig. \ref{gating_all}), indicating that this behavior is not specific to a particular gating configuration or charge-doping method.

\subsection*{Breakdown of the anti-Stokes/Stokes thermometry}

The implications of the gate-dependent anti-Stokes response were next examined using Raman thermometry based on the Bose-Einstein phonon population. Fig.~\ref{fig3}a shows the apparent temperature extracted from the anti-Stokes/Stokes areal intensity ratio using Eq. \ref{eq1},~\cite{mcgrane2014,Cardona1982} plotted against $|E_\textrm{F}|$:

\begin{equation}
\frac{I_{\mathrm{aS}}}{I_{\mathrm{S}}} = \left(\frac{\omega_{\mathrm{L}}+\omega_{\mathrm{G}}}{\omega_{\mathrm{L}}-\omega_{\mathrm{G}}}\right)^3 \exp\left(-\frac{\hbar\omega_{\mathrm{G}}}{k_{\mathrm{B}} T}\right).
\label{eq1}
\end{equation}

Here, $\omega_{\mathrm{L}}$ is the laser excitation frequency, $\hbar\omega_{\mathrm{G}}$ the G-phonon energy, and $k_{\mathrm{B}}$ the Boltzmann constant. The $(\omega_{\mathrm{L}}+\omega_{\mathrm{G}})^3/(\omega_{\mathrm{L}}-\omega_{\mathrm{G}})^3$ term reflects the ratio of scattered to incident photon number and should be appropriate when the instrumental response function is known over the measured spectral range.\cite{mcgrane2014} In our measurements, the Stokes and anti-Stokes intensities were additionally corrected for the wavelength-dependent instrumental response using the calibration-lamp response of the Raman system. The resulting $I_{\mathrm{aS}}/I_{\mathrm{S}}$ values therefore represent background-subtracted, instrument-response-corrected intensity ratios, without empirical normalization to room temperature. In the literature, the thermometry expression is often supplemented or replaced by an empirical parameter $C$, which can account for instrumental response and other sample-specific effects.\cite{alaferdov2020} Such normalization effectively forces the measured ratio toward the expected temperature and may therefore obscure intrinsic deviations from the equilibrium thermometric response.

In our experiment, the temperature ($T$) inferred from the graphene G mode (calculated directly from the instrument-response-corrected $I_{\mathrm{aS}}/I_{\mathrm{S}}$ ratio using Eq. \ref{eq1}, corresponding to $C=1$ with no additional empirical normalization) decreases monotonically with increasing doping, from $\approx590$~K near charge neutrality to $\approx450$~K at higher carrier densities. This apparently high graphene temperature reflects an anomalously large anti-Stokes intensity relative to the Stokes counterpart, rather than a genuine elevation of the lattice temperature. Normalization of such $I_{\mathrm{aS}}/I_{\mathrm{S}}$ ratios to room temperature would yield the $C$ factor ranging between $\approx$0.5 and $\approx$0.7, which partially explains the large span of $C$ values found in graphene-thermometry-related literature \cite{sullivan2017,berciaud2010,sun2017,faugeras2010,chae2010}. By contrast, the reference temperature, which we extracted from the substrate Si mode at $\mathrm{520\ cm^{-1}}$/$\mathrm{-520\ cm^{-1}}$ using the same instrument-response correction and $C=1$, remains close to room temperature throughout the same gating range. Since the Si substrate is probed through the same optical path and electrolyte layer as the graphene, any gate-dependent change in electrolyte absorption or the formation of absorbing interfacial layers would be expected to affect both signals comparably. The absence of a similar variation in the Si reference strongly argues against changes in electrolyte absorption, absorbing interfacial layers, or other optical-path effects as the origin of the observed graphene response. This discrepancy shows that the standard thermometric interpretation of $I_{\mathrm{aS}}/I_{\mathrm{S}}$ does not provide a self-consistent description of the graphene response under Fermi-level tuning.

\subsection*{Power dependence}

\begin{figure}[h!]
    \centering
    \includegraphics[width=0.5\linewidth]{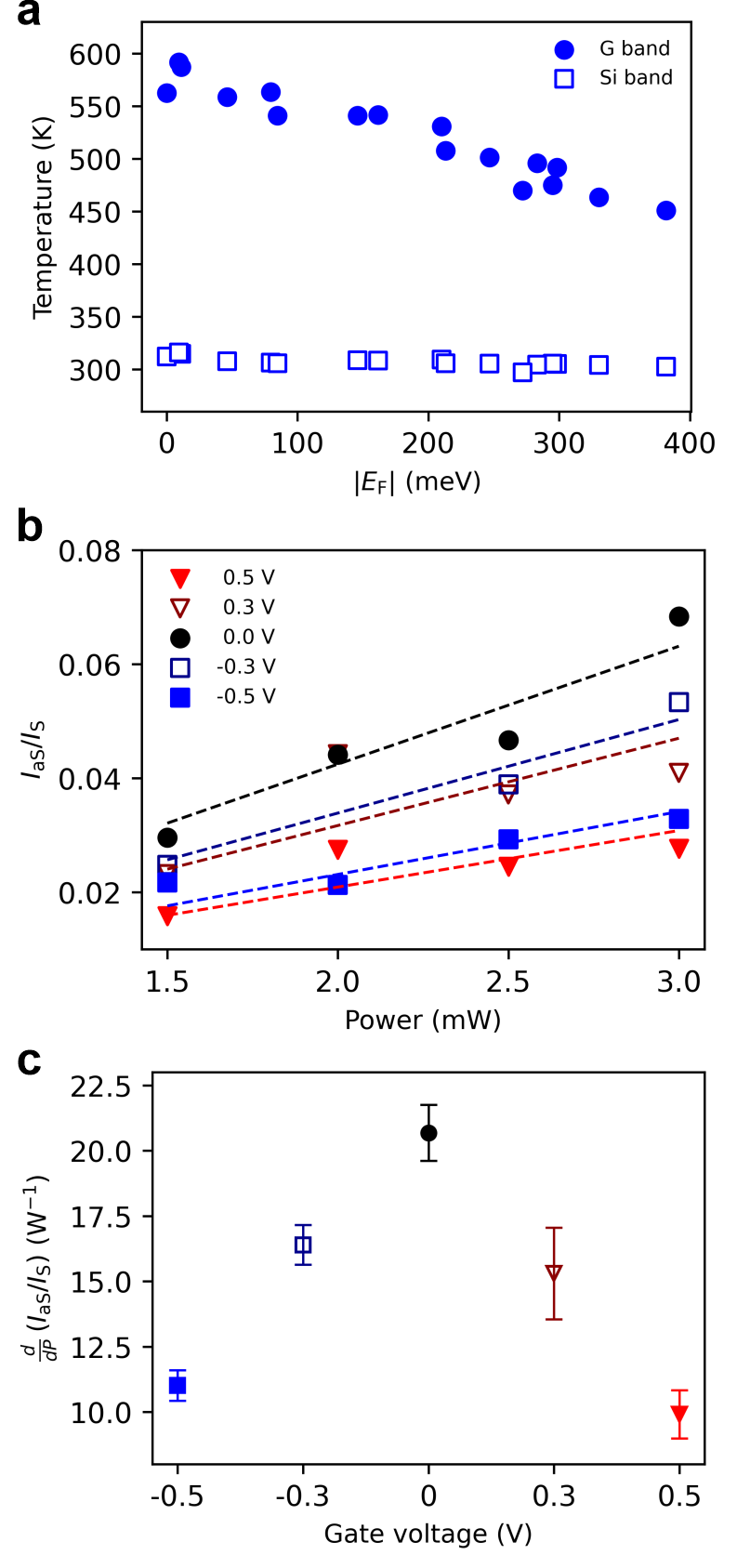}
    \caption{\textbf{Breakdown of the Raman thermometry and gate-dependent power response of the anti-Stokes/Stokes ratio.} \textbf{a,} Temperature extracted from  $I_{\mathrm{aS}}/I_{\mathrm{S}}$ for the graphene G band and the reference Si band at $\mathrm{520~cm^{-1}}$ using Eq. \ref{eq1} plotted against the absolute Fermi level shift (laser excitation wavelength of 633~nm, 1.5~mW power at the sample, 3.8~M LiClO\textsubscript{4} aq. electrolyte). \textbf{b,} Power dependence of $I_{\mathrm{aS}}/I_{\mathrm{S}}$ for different gate voltages. \textbf{c,} Slope of the $I_{\mathrm{aS}}/I_{\mathrm{S}}$-power dependence, fitted in \textbf{b} (dashed lines), as a function of gate voltage.}
    \label{fig3}
\end{figure}

To further investigate the origin of the Raman thermometry breakdown, the anti-Stokes/Stokes areal intensity ratio was measured as a function of the excitation power for several gate voltages. In all cases, $I_{\mathrm{aS}}/I_{\mathrm{S}}$ increases approximately linearly over the investigated power range, with the largest increase observed near the Dirac point and progressively weaker changes away from it (Fig.~\ref{fig3}b). Such behavior may arise from power-dependent nonequilibrium phonon populations and is also compatible with a correlated Stokes–anti-Stokes contribution, for which $I_{\mathrm{S}}\propto P$ and the correlated anti-Stokes contribution scales approximately as $I_{\mathrm{aS}}\propto P^2$, yielding $I_{\mathrm{aS}}/I_{\mathrm{S}}\propto P$.~\cite{parra2016} The observed power dependence alone therefore does not distinguish between these contributions; its pronounced gate dependence instead demonstrates that the power-induced anti-Stokes enhancement is sensitive to the electronic state of graphene. To compare the effective power response across different gate voltages, the linear fits in Fig.~\ref{fig3}b were constrained to the shared $P\rightarrow 0$ thermal-equilibrium limit $I_{\mathrm{aS}}/I_{\mathrm{S}} = 0.0011$, calculated from Eq. \ref{eq1} at 300~K for 633~nm laser excitation. The extracted effective slopes, summarized in Fig.~\ref{fig3}c, exhibit a symmetric decrease as the Fermi level shifts away from the Dirac point. These measurements show that both the magnitude of the anti-Stokes response and its power scaling are strongly controlled by carrier density, in contrast to the nearly gate-independent Stokes intensity.

\subsection*{Continuous evolution of anti-Stokes/Stokes ratio with gating}

\begin{figure}[h!]
    \centering
    \includegraphics[width=0.5\linewidth]{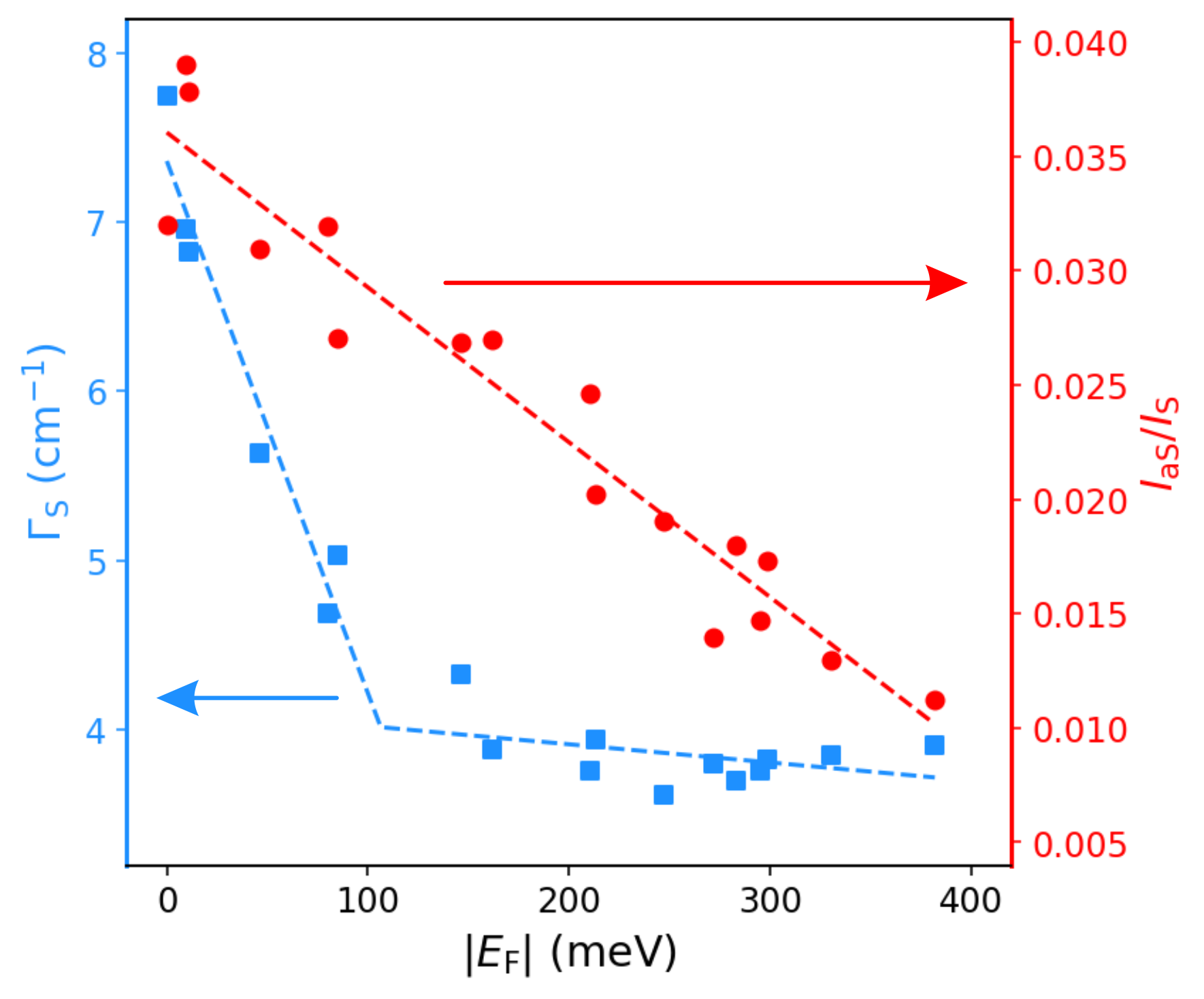}
    \caption{\textbf{Evolution of the G-band anti-Stokes/Stokes intensity ratio and linewidth with Fermi-level shift.} The G-band Stokes linewidth $\Gamma_\mathrm{S}$ (blue squares, left axis) and the anti-Stokes-to-Stokes integrated intensity ratio $I_{\mathrm{aS}}/I_{\mathrm{S}}$ (red circles, right axis) are plotted as a function of the absolute Fermi-level shift $|E_\mathrm{F}|$. Dashed lines represent the corresponding linear and bilinear fits to the experimental data. (The data plotted here correspond to the sample shown in Fig.~\ref{fig2}; the evolution of $\Gamma_\mathrm{S}$ and $I_{\mathrm{aS}}/I_{\mathrm{S}}$ with $|E_\mathrm{F}|$ for all other liquid-gated samples is shown in Supporting Fig. \ref{bilinear_fitting_sec}).}
    \label{fig_IaSIS_Gamma}
\end{figure}

For each device, $I_{\mathrm{aS}}/I_{\mathrm{S}}$ and the Stokes G‑mode linewidth, $\Gamma_\textrm{S}$, were analyzed as functions of $|E_\textrm{F}|$. A representative dataset is shown in Fig.~\ref{fig_IaSIS_Gamma}, where $I_{\mathrm{aS}}/I_{\mathrm{S}}$ exhibits a nearly linear decrease with $|E_\textrm{F}|$ over the entire probed range, whereas $\Gamma_\textrm{S}(|E_\textrm{F}|)$ is well described by a bilinear dependence with a crossover at $|E_\textrm{F}|\approx 100$~meV, i.e., slightly above $\hbar\omega_{\mathrm{G}}$/2. We note that at room temperature, the sharp Fermi-level threshold expected at zero temperature is broadened by the Fermi–Dirac distribution; the bilinear fit is therefore used here as a simplified empirical description of the observed crossover. Similar bilinear behavior of $\Gamma_\textrm{S}$ is observed for all devices (Supporting Fig. \ref{bilinear_fitting_sec}), consistent with the established removal of the Kohn anomaly once $|E_\textrm{F}|>\hbar\omega_{\mathrm{G}}/2$.~\cite{lazzeri2006} In contrast, $I_{\mathrm{aS}}/I_{\mathrm{S}}$ is well described by a single linear trend over the accessible doping range, without a discernible change of slope at $|E_\textrm{F}|\approx \hbar\omega_{\mathrm{G}}/2$.

\section*{Discussion}\label{discussion}

Our results reveal a sizable deviation from the Raman thermometry interpretation of the anti-Stokes/Stokes intensity ratio, in which $I_{\mathrm{aS}}/I_{\mathrm{S}}$ is determined by the thermal phonon population and directly reflects the lattice temperature. However, the strong and symmetric dependence of the G-mode-derived apparent temperature on carrier density (Fig.~\ref{fig3}a), absent in the Si mode reference, demonstrates that this description breaks down for graphene, with the discrepancy becoming particularly evident under Fermi-level tuning. A purely thermal interpretation would require a reduction of the lattice temperature by more than 100~K upon electrostatic or chemical doping under identical illumination conditions. No known mechanism can account for such efficient lattice cooling during gating under continuous-wave (CW) optical excitation. Furthermore, the substrate Si mode, which is sensitive to the same local thermal environment, remains close to room temperature, arguing against substantial gate-dependent substrate heating or an optical-path artifact. A related departure from conventional anti-Stokes Raman thermometry was reported by Ferrante et al. under picosecond excitation of graphene.~\cite{Ferrante2018} In such a strongly non-equilibrium regime, the authors argued that the anti-Stokes/Stokes ratio may reflect not only the phonon population but also wavelength-dependent resonant Raman efficiencies and possible correlated Stokes–anti-Stokes processes.~\cite{Ferrante2018} Our measurements extend this phenomenology to CW excitation and reveal that the anomalous anti-Stokes response can be controlled by the Fermi level, while the nearly unchanged Stokes response and the substrate reference disfavor a purely thermal interpretation. The observed behavior therefore requires a non-thermal, electronically sensitive contribution to the anti-Stokes response.

A key insight into the nature of this contribution is provided by comparing the gate dependence of $I_{\mathrm{aS}}/I_{\mathrm{S}}$ with that of $\Gamma_\textrm{S}$, which serves as an independent probe of the electron–phonon coupling strength. As shown in Fig.~\ref{fig_IaSIS_Gamma}, $\Gamma_\textrm{S}$ follows a bilinear dependence on $|E_\textrm{F}|$, with a clear crossover near $|E_\textrm{F}|\approx \hbar\omega_{\mathrm{G}}/2 \approx 100$~meV, consistent with the well-established removal of the Kohn anomaly once Pauli blocking suppresses the decay of G-mode phonons into electron–hole pairs.~\cite{lazzeri2006} In contrast, $I_{\mathrm{aS}}/I_{\mathrm{S}}(|E_\textrm{F}|)$ varies smoothly and approximately linearly across the entire probed doping range, without a discernible change of slope at the Kohn-anomaly threshold. This decoupling of the anti-Stokes response from the linewidth behavior shows that $I_{\mathrm{aS}}/I_{\mathrm{S}}$ is not simply tracking the local electron–phonon matrix element at the G-phonon energy. Furthermore, within the two-photon formalism of the Stokes–anti-Stokes scattering framework of Guim\~{a}raes et al., the observed evolution of the Raman linewidth alone would be expected to produce the opposite trend in the anti-Stokes response.~\cite{guimaraes2020} The linewidth variation therefore cannot account for the gate-dependent suppression of $I_{\mathrm{aS}}/I_{\mathrm{S}}$, further pointing to a role of the electronic system beyond the linewidth renormalization.

Two possible contributions may account for this behavior. First, increasing carrier density raises the electronic heat capacity, as emphasized by Bistritzer and MacDonald in the context of electronic cooling in graphene's Dirac cone,~\cite{bistritzer2009} and enhances electron–lattice energy transfer.~\cite{graham2013} Under fixed CW optical excitation, this progressively lowers the steady-state hot-electron and hot-optical-phonon temperatures, reducing the non-equilibrium phonon population that drives the anti-Stokes signal above its thermal baseline — an effect directly evidenced by the doping-dependent hot-carrier temperatures and cooling dynamics observed in time-resolved-angle-resolved photoemission spectroscopy and photocurrent experiments.~\cite{weigl2026,johannsen2015,graham2013} Second, the anti-Stokes intensity may contain a contribution from electronically mediated Stokes–anti-Stokes coupling, whose efficiency depends on the electronic phase space available near the Fermi level. Both contributions are expected to weaken with increasing carrier density and therefore act in the same direction.

\begin{figure}[h!]
    \centering
    \includegraphics[width=0.75\linewidth]{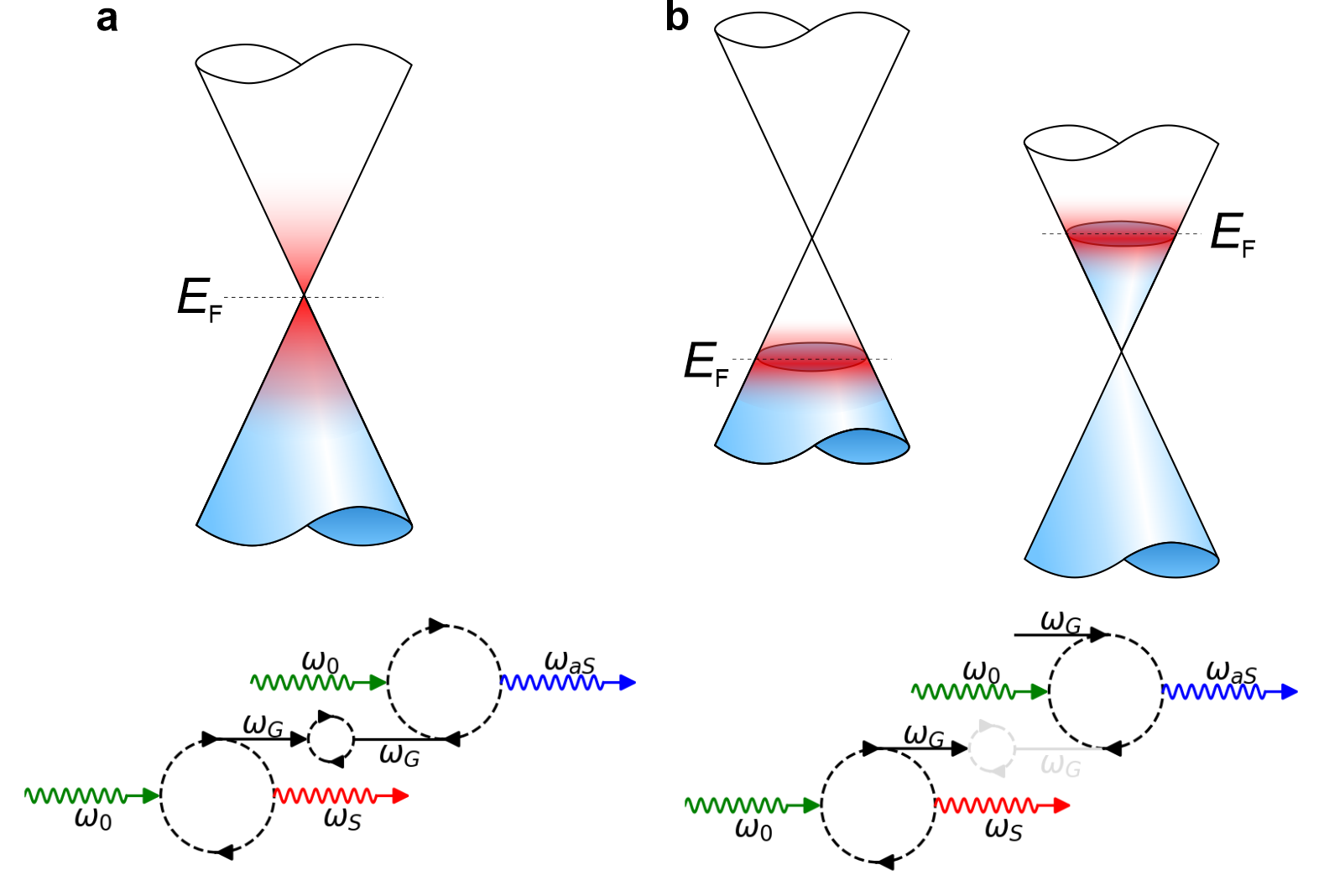}
    \caption{
    \textbf{Electronically mediated Stokes–anti-Stokes coupling and its suppression with doping.} \textbf{a,} Dirac cone at charge neutrality ($E_\mathrm{F}\approx 0$), with the red-shaded region indicating the electronic phase space available for coupling between the Stokes and anti-Stokes channels, and the corresponding Feynman diagram illustrating the correlated scattering pathway mediated by this electronic continuum. \textbf{b,}  Dirac cone with $E_\mathrm{F}$ shifted away from charge neutrality: the reduced overlap between occupied and empty electronic states (blue-shaded region) restricts the electronic phase space available for the proposed coupling, as illustrated in the corresponding Feynman diagram.}
    \label{fig5}
\end{figure}

We use the term Stokes–anti-Stokes coupling here in the broader sense of an electronically mediated link between the two scattering channels, rather than as evidence of correlated photon-pair generation, which would require photon-correlation measurements. In the mechanism previously discussed for graphene and diamond, a phonon created in the Stokes process can contribute to subsequent anti-Stokes scattering, enhancing $I_{\mathrm{aS}}/I_{\mathrm{S}}$ beyond its thermal equilibrium value.~\cite{jorio2015,kasperczyk2015}  The corresponding electronic phase space is illustrated schematically in Fig. \ref{fig5}, where the red-shaded region in Fig. \ref{fig5}a reflects the maximal electronic continuum available for Stokes–anti-Stokes coupling at charge neutrality and becomes restricted as $E_\mathrm{F}$ moves away from the Dirac point in Fig. \ref{fig5}b. Unlike the comparatively sharp change in the G-mode linewidth near $E_\mathrm{F}\approx \hbar\omega_{\mathrm{G}}/2$ (Fig. \ref{fig_IaSIS_Gamma}, blue), the phase space for electronic scattering events involving photon and phonon exchange extends over a broader energy window and may therefore decrease smoothly across the doping range investigated here (Fig. \ref{fig_IaSIS_Gamma}, red). This provides a possible explanation for the nearly continuous suppression of $I_{\mathrm{aS}}/I_{\mathrm{S}}$ with increasing $E_\mathrm{F}$.

Direct evidence for correlated Stokes–anti-Stokes scattering has been obtained in diamond and in specially engineered twisted bilayer graphene under resonantly enhanced conditions.~\cite{kasperczyk2015,parra2016} By contrast, for single-layer graphene under CW excitation, the power dependence of $I_{\mathrm{aS}}/I_{\mathrm{S}}$ has previously been attributed primarily to laser-induced heating.~\cite{jorio2015} The present measurements do not distinguish a correlated scattering contribution from a doping-dependent nonequilibrium phonon population; they instead show that the anti-Stokes response contains an electronically sensitive component that is strongest near charge neutrality and weakens with increasing doping.

This interpretation is not contradicted by time-resolved pump–probe studies of gated graphene field-effect transistors, which reported a G-phonon population lifetime independent of $E_\textrm{F}$ and attributed it to a cancellation of Landau-damping channels once carriers and phonons become mutually thermalized under pulsed excitation.~\cite{Wu2012} That measurement probes the intrinsic phonon lifetime after transient thermalization under high-fluence pulsed excitation, whereas our CW experiment probes a steady-state anti-Stokes response determined by the continuous balance between phonon generation and dissipation. The latter can remain sensitive to the Fermi-level-dependent electronic phase space even when the intrinsic phonon lifetime is not.

Notably, the anomalously short optical-phonon lifetimes reported for monolayer and bilayer graphene — attributed to strong coupling between the G phonon and the photoexcited electronic continuum~\cite{kang2010} — were later invoked as an indirect support for a correlated Stokes–anti-Stokes channel in twisted bilayer graphene.~\cite{parra2016} The doping dependence observed here is consistent with the proposed link between the electronic system and the Stokes–anti-Stokes response, with the electronically mediated contribution being strongest near charge neutrality and weakening with increasing doping. By continuously tuning $E_\mathrm{F}$, our measurements provide doping-resolved evidence that such a contribution may also be present in ordinary monolayer graphene, although its magnitude cannot be separated from that of the nonequilibrium phonon population in the present experiment. 

\section*{Conclusions}

In summary, gate-tunable Stokes–anti-Stokes Raman spectroscopy provides a sensitive, all-optical probe of the non-equilibrium electronic and phonon states in graphene. The correlated evolution of $I_{\mathrm{aS}}/I_{\mathrm{S}}$ with Fermi energy and laser power is consistent with contributions from both doping-dependent non-equilibrium carrier/phonon populations and an additional electronically mediated Raman response. Although the present measurements do not separate these contributions, they demonstrate that the anti-Stokes response of monolayer graphene contains a strong electronic component that is maximal near charge neutrality and progressively suppressed with increasing $|E_\mathrm{F}|$. The latter mechanism has so far only been directly confirmed in resonantly enhanced twisted bilayer graphene and hinted at by the anomalously short optical-phonon lifetimes reported in non-resonant monolayer and bilayer graphene. Our work indicates that such a contribution may also be present and continuously tunable in ordinary monolayer graphene. Together, our results demonstrate that the anti-Stokes Raman response provides a sensitive probe of Fermi-level-dependent electronic and phononic dynamics in graphene, complementary to transport- and photoluminescence-based approaches. This measurement was made possible by the localized microdroplet platform, which uniquely combines the high carrier-density range of electrolyte gating with an electrolyte layer thin enough to transmit the weak anti-Stokes signal — a combination that dielectric gating, thick-electrolyte cells, chemical doping, and inverted-microscope geometries each fail to provide simultaneously. Future work employing pulsed excitation and photon-correlation detection would allow the thermometric and correlation-driven contributions to be cleanly separated, which is particularly pertinent to twisted bilayer graphene near the magic angle, where both the electronic heat capacity and the electron–phonon coupling are strongly renormalized. More broadly, our results reinforce the growing recognition that anti-Stokes Raman intensities in low-dimensional materials can encode non-equilibrium electronic and interfacial physics beyond simple lattice heating, as recently highlighted in van der Waals heterostructures exhibiting non-thermal optical cooling.

\section*{Supporting Information}
Experimental methods, Raman calibration and fitting procedures, sample characterization, complementary dielectric- and chemical-gating measurements, and additional gate-dependent Raman datasets.

\begin{acknowledgement}

M.J. and O.F. thank the Czech Science Foundation (project No. 23-06174K) for financial support. M.V. acknowledges the support of the Lumina Quaeruntur fellowship No. LQ200402201 by the Czech Academy of Sciences. A.J. acknowledges Conselho Nacional de Desenvolvimento Científico e Tecnológico (CNPq, Grant No. 421469/2023-4) and Fundação de Amparo à Pesquisa do Estado de Minas Gerais (FAPEMIG, Grants No. APQ-04852-23 and No. RED 00081-23). R.S. acknowledges the Yushan Fellow Program by the MOE in Taiwan. This work was also supported by the Ministry of Education, Youth, and Sports of the Czech Republic, Project No. CZ.02.01.01/00/22\_008/0004558, co-funded by the European Union, and the CzechNanoLab Research Infrastructure, supported by the Ministry of Education, Youth, and Sports of the Czech Republic (LM2023051). Support of the Nanofabrication Facilities Rossendorf (NanoFaRo) at the IBC is gratefully acknowledged.

\end{acknowledgement}

\section*{Data availability}

The data underlying this study are available from the HeyRACK repository at [persistent link to data repository TBC]

\section*{Competing interest}

The authors declare no competing interests.

\section*{Authors contribution}
M.J. performed the experiments, prepared samples, analyzed the data, prepared the figures, wrote the original draft of the manuscript and contributed to the manuscript revision. L.P. performed the experiments and contributed to the manuscript revision. T.V. contributed to the conceptualization of the study and to the manuscript revision. C.F. contributed to sample preparation. M.K. contributed to the funding acquisition and manuscript revision. M.V. contributed to the conceptualization of the study, funding acquisition, and manuscript revision. R.S. and A.J. contributed to scientific discussions and manuscript revision. O.F. conceived and supervised the study, acquired funding, coordinated the project, and contributed to the manuscript revision. All authors discussed the results and approved the final manuscript.

\clearpage

\clearpage
\bibliography{SaS_bibliography}

@article{alaferdov2020,
   author = {Alaferdov, A. V. and Savu, R. and Fantini, C. and Cancado, L. G. and Moshkalev, S. A.},
   title = {Raman spectra of multilayer graphene under high temperatures},
   journal = {J. Phys.: Condens. Matter},
   volume = {32},
   number = {38},
   pages = {385704},
   ISSN = {1361-648X-0953-8984},
   DOI = {https://doi.org/10.1088/1361-648x/ab95ce},
   year = {2020},
   type = {Journal Article}
}

@article{ando2006,
  title={Anomaly of optical phonon in monolayer graphene},
  author={Ando, Tsuneya},
  journal={J. Phys. Soc. Jpn.},
  volume={75},
  number={12},
  pages={124701},
  year={2006},
  doi={https://doi.org/10.1143/jpsj.75.124701},
  publisher={The Physical Society of Japan}
}

@article{balandin2011,
   author = {Balandin, Alexander A.},
   title = {Thermal Properties of Graphene and Nanostructured Carbon Materials},
   journal = {Nat. Mater.},
   volume = {10},
   number = {8},
   pages = {569-581},
   ISSN = {1476-1122},
   doi = {https://doi.org/10.1038/nmat3064},
   url = {http://dx.doi.org/10.1038/nmat3064},
   year = {2011},
   type = {Journal Article}
}

@article{berciaud2010,
  title={Electron and optical phonon temperatures in electrically biased graphene},
  author={Berciaud, St{\'e}phane and Han, Melinda Y. and Mak, Kin Fai and Brus, Louis E. and Kim, Philip and Heinz, Tony F.},
  journal={Phys. Rev. Lett.},
  volume={104},
  number={22},
  pages={227401},
  year={2010},
  doi={https://doi.org/10.1103/physrevlett.104.227401},
  publisher={APS}
  
}

@article{bistritzer2009,
   author = {Bistritzer, R. and MacDonald, A. H.},
   title = {Electronic Cooling in Graphene},
   journal = {Phys. Rev. Lett.},
   volume = {102},
   number = {20},
   pages = {206410},
   DOI = {https://doi.org/10.1103/physrevlett.102.206410},
   url = {https://link.aps.org/doi/10.1103/PhysRevLett.102.206410},
   year = {2009},
   type = {Journal Article}
}

@article{breusing2011,
   author = {Breusing, M. and Kuehn, S. and Winzer, T. and Malić, E. and Milde, F. and Severin, N. and Rabe, J. P. and Ropers, C. and Knorr, A. and Elsaesser, T.},
   title = {Ultrafast Nonequilibrium Carrier Dynamics in a Single Graphene Layer},
   journal = {Phys. Rev. B},
   volume = {83},
   number = {15},
   pages = {153410},
   DOI = {10.1103/PhysRevB.83.153410},
   url = {https://link.aps.org/doi/10.1103/PhysRevB.83.153410},
   year = {2011},
   type = {Journal Article}
}

@inbook{Cardona1982,
   author = {Cardona, M.},
   title = {Resonance Phenomena},
   booktitle = {Light Scattering in Solids Ii: Basic Concepts and Instrumentation},
   editor = {Cardona, Manuel and Güntherodt, Gernot},
   publisher = {Springer Berlin Heidelberg},
   address = {Berlin, Heidelberg},
   pages = {19-178},
   ISBN = {978-3-540-39075-6},
   DOI = {https://doi.org/10.1007/3-540-11380-0\_14},
   year = {1982},
   type = {Book Section}
}

@article{chae2010,
  title={Hot phonons in an electrically biased graphene constriction},
  author={Chae, Dong-Hun and Krauss, Benjamin and Von Klitzing, Klaus and Smet, Jurgen H},
  journal={Nano Lett.},
  volume={10},
  number={2},
  pages={466--471},
  year={2010},
  doi = {https://doi.org/10.1021/nl903167f},
  publisher={ACS Publications}
}

@article{chen2011,
  title={Controlling inelastic light scattering quantum pathways in graphene},
  author = {Chen,  Chi-Fan and Park,  Cheol-Hwan and Boudouris,  Bryan W. and Horng,  Jason and Geng,  Baisong and Girit,  Caglar and Zettl,  Alex and Crommie,  Michael F. and Segalman,  Rachel A. and Louie,  Steven G. and Wang,  Feng},
  journal={Nature},
  volume={471},
  number={7340},
  pages={617--620},
  year={2011},
  doi={https://doi.org/10.1038/nature09866},
  publisher={Nature Publishing Group UK London}
  
}

@article{das2008,
   author = {Das, A. and Pisana, S. and Chakraborty, B. and Piscanec, S. and Saha, S. K. and Waghmare, U. V. and Novoselov, K. S. and Krishnamurthy, H. R. and Geim, A. K. and Ferrari, A. C. and Sood, A. K.},
   title = {Monitoring dopants by Raman scattering in an electrochemically top-gated graphene transistor},
   journal = {Nat. Nanotechnol.},
   volume = {3},
   number = {4},
   pages= {210–215},
   ISSN = {1748-3395},
   DOI = {https://doi.org/10.1038/nnano.2008.67},
   url = {https://www.nature.com/articles/nnano.2008.67},
   year = {2008},
   type = {Journal Article}
}

@article{faugeras2010,
  title={Thermal conductivity of graphene in corbino membrane geometry},
  author={Faugeras, Clement and Faugeras, Blaise and Orlita, Milan and Potemski, Marek and Nair, Rahul R. and Geim, A. K.},
  journal={ACS Nano},
  volume={4},
  number={4},
  pages={1889--1892},
  year={2010},
  publisher={ACS Publications},
  doi={https://doi.org/10.1021/nn9016229}
}

@article{Ferrante2018,
   author = {Ferrante, C. and Virga, A. and Benfatto, L. and Martinati, M. and De Fazio, D. and Sassi, U. and Fasolato, C. and Ott, A. K. and Postorino, P. and Yoon, D. and Cerullo, G. and Mauri, F. and Ferrari, A. C. and Scopigno, T.},
   title = {Raman Spectroscopy of Graphene under Ultrafast Laser Excitation},
   journal = {Nat. Commun.},
   volume = {9},
   number = {1},
   pages = {308},
   ISSN = {2041-1723},
   DOI = {10.1038/s41467-017-02508-x},
   url = {https://doi.org/10.1038/s41467-017-02508-x
https://www.nature.com/articles/s41467-017-02508-x.pdf},
   year = {2018},
   type = {Journal Article}
}

@article{ferrari2006,
   author = {Ferrari, A. C. and Meyer, J. C. and Scardaci, V. and Casiraghi, C. and Lazzeri, M. and Mauri, F. and Piscanec, S. and Jiang, D. and Novoselov, K. S. and Roth, S. and Geim, A. K.},
   title = {Raman spectrum of graphene and graphene layers},
   journal = {Phys. Rev. Lett.},
   volume = {97},
   number = {18},
   pages = {187401},
   ISSN = {0031-9007, 0031-9007},
   DOI = {https://doi.org/10.1103/PhysRevLett.97.187401},
   url = {https://www.ncbi.nlm.nih.gov/pubmed/17155573},
   year = {2006},
   type = {Journal Article}
}

@article{graham2013,
   author = {Graham, Matt W. and Shi, Su-Fei and Ralph, Daniel C. and Park, Jiwoong and McEuen, Paul L.},
   title = {Photocurrent Measurements of Supercollision Cooling in Graphene},
   journal = {Nat. Phys.},
   volume = {9},
   number = {2},
   pages = {103-108},
   ISSN = {1745-2481},
   DOI = {https://doi.org/10.1038/nphys2493},
   url = {https://doi.org/10.1038/nphys2493},
   year = {2013},
   type = {Journal Article}
}

@article{guimaraes2020,
  title={Stokes--anti-{Stokes} light-scattering process: A photon-wave-function approach},
  author={Guimar{\~a}es, Alfredo Vin{\'\i}cius Andrade and Santos, Marcelo F and Jorio, A and Monken, Carlos Henrique},
  journal={Phys. Rev. A},
  volume={102},
  number={3},
  pages={033719},
  year={2020},
  publisher={APS},
  DOI = {https://doi.org/10.1103/PhysRevA.102.033719}
}

@article{hasdeo2014,
  title={Breit-{W}igner-{F}ano line shapes in {R}aman spectra of graphene},
  author={Hasdeo, Eddwi H. and Nugraha, Ahmad R. T. and Dresselhaus, Mildred S. and Saito, Riichiro},
  journal={Phys. Rev. B},
  volume={90},
  number={24},
  pages={245140},
  year={2014},
  doi={https://doi.org/10.1103/PhysRevB.90.245140},
  publisher={APS}
}

@article{johannsen2015,
   author = {Johannsen, Jens Christian and Ulstrup, Søren and Crepaldi, Alberto and Cilento, Federico and Zacchigna, Michele and Miwa, Jill A. and Cacho, Cephise and Chapman, Richard T. and Springate, Emma and Fromm, Felix and Raidel, Christian and Seyller, Thomas and King, Phil D. C. and Parmigiani, Fulvio and Grioni, Marco and Hofmann, Philip},
   title = {Tunable Carrier Multiplication and Cooling in Graphene},
   journal = {Nano Lett.},
   volume = {15},
   number = {1},
   pages = {326-331},
   ISSN = {1530-6984},
   DOI = {https://doi.org/10.1021/nl503614v},
   url = {https://doi.org/10.1021/nl503614v},
   year = {2015},
   type = {Journal Article}
}

@article{jorio2015,
   author = {Jorio, Ado and Kasperczyk, Mark and Clark, Nick and Neu, Elke and Maletinsky, Patrick and Vijayaraghavan, Aravind and Novotny, Lukas},
   title = {Stokes and anti-{S}tokes {R}aman spectra of the high-energy {C-C} stretching modes in graphene and diamond},
   journal = {Phys. Status Solidi B},
   volume = {252},
   number = {11},
   pages = {2380-2384},
   ISSN = {0370-1972},
   DOI = {https://doi.org/10.1002/pssb.201552224},
   year = {2015},
   type = {Journal Article}
}

@article{kang2010,
   author = {Kang, Kwangu and Abdula, Daner and Cahill, David G. and Shim, Moonsub},
   title = {Lifetimes of optical phonons in graphene and graphite by time-resolved incoherent anti-{S}tokes {R}aman scattering},
   journal = {Phys. Rev. B},
   volume = {81},
   number = {16},
   pages = {165405},
   ISSN = {1098-0121},
   DOI = {https://doi.org/10.1103/physrevb.81.165405},
   year = {2010},
   type = {Journal Article}
}

@article{kasperczyk2015,
  title={Stokes--anti-{S}tokes correlations in diamond},
  author={Kasperczyk, Mark and Jorio, Ado and Neu, Elke and Maletinsky, Patrick and Novotny, Lukas},
  journal={Opt. Lett.},
  volume={40},
  number={10},
  pages={2393--2396},
  year={2015},
   doi={https://doi.org/10.1364/OL.40.002393},
  publisher={Optical Society of America} 
}

@article{lazzeri2006,
  title={Nonadiabatic {K}ohn anomaly in a doped graphene monolayer},
  author = {Lazzeri, M. and Mauri, F.},
  journal = {Phys. Rev. Lett.},
  volume = {97},
  issue = {26},
  pages = {266407},
  numpages = {4},
  year = {2006},
  doi = {https://doi.org/10.1103/PhysRevLett.97.266407},
  url = {https://link.aps.org/doi/10.1103/PhysRevLett.97.266407},
  publisher = {American Physical Society}
 
}

@article{lin2015,
   author = {Lin, Kuan-Chun and Li, Ming-Yang and Ling, D. C. and Chi, C. C. and Chen, Jeng-Chung},
   title = {Evolution of Hot Carrier Dynamics in Graphene with the {F}ermi Level Tuned across the Dirac Point},
   journal = {Phys. Rev. B},
   volume = {91},
   number = {12},
   pages = {125440},
   DOI = {https://doi.org/10.1103/PhysRevB.91.125440},
   url = {https://link.aps.org/doi/10.1103/PhysRevB.91.125440},
   year = {2015},
   type = {Journal Article}
}

@article{lin2026,
   author = {Lin, Jiamin and Xiang, Baixu and Liu, Renguang and Ling, Jinyang and Wang, Gang and Zhang, Le and Li, Li and Li, Hua and Zhang, Dongxu and Duan, Zhexing and Zhang, Qi and Wan, Changjin and Wang, Wei and Wang, Xingzhi and Lin, Junhao and Gao, Huajian and Xiong, Qihua and Xu, Weigao},
   title = {Optical Cooling by Interfacial Charge Transfer in {2D} Heterostructures},
   journal = {Nature},
   volume = {655},
   number = {8122},
   pages = {342-349},
   ISSN = {1476-4687},
   DOI = {https://doi.org/10.1038/s41586-026-10662-w},
   url = {https://doi.org/10.1038/s41586-026-10662-w},
   year = {2026},
   type = {Journal Article}
}

@article{malard2009,
   author = {Malard, L. M. and Pimenta, M. A. and Dresselhaus, G. and Dresselhaus, M. S.},
   title = {Raman spectroscopy in graphene},
   journal = {Phys. Rep.},
   volume = {473},
   number = {5-6},
   pages = {51-87},
   ISSN = {03701573},
   DOI = {https://doi.org/10.1016/j.physrep.2009.02.003},
   year = {2009},
   type = {Journal Article}
}

@article{mcgrane2014,
   author = {McGrane, Shawn D. and Moore, David S. and Goodwin, Peter M. and Dattelbaum, Dana M.},
   title = {Quantitative Tradeoffs between Spatial, Temporal, and Thermometric Resolution of Nonresonant {R}aman Thermometry for Dynamic Experiments},
   journal = {Appl. Spectrosc.},
   volume = {68},
   number = {11},
   pages = {1279-1288},
   DOI = {https://doi.org/10.1366/14-07503},
   url = {https://journals.sagepub.com/doi/abs/10.1366/14-07503},
   year = {2014},
   type = {Journal Article}
}

@article{nadas2023,
  title={Spatially coherent tip-enhanced {R}aman spectroscopy measurements of electron--phonon interaction in a graphene device},
  author={Nadas, Rafael Battistella and Gadelha, Andreij C and Barbosa, Tiago C and Rabelo, Cassiano and de Lourenco e Vasconcelos, Thiago and Monken, Vitor and Portes, Ary VR and Watanabe, Kenji and Taniguchi, Takashi and Ramirez, Jhonattan C and others},
  journal={Nano Lett.},
  volume={23},
  number={19},
  pages={8827--8832},
  year={2023},
  publisher={ACS Publications},
  DOI = {https://doi.org/10.1021/acs.nanolett.3c00851}
}

@article{neto2009,
  title={The electronic properties of graphene},
  author={Castro Neto, Antonio H. and Guinea, Francisco and Peres, Nuno M. R. and Novoselov, Kostya S. and Geim, Andre K.},
  journal={Rev. Mod. Phys.},
  volume={81},
  number={1},
  pages={109--162},
  year={2009},
  doi={https://doi.org/10.1103/RevModPhys.81.109},
  publisher={APS}
}

@article{neto2007,
  title={Electron-phonon coupling and {R}aman spectroscopy in graphene},
  author={Castro Neto, AH and Guinea, Francisco},
  journal={Phys. Rev. B},
  volume={75},
  number={4},
  pages={045404},
  year={2007},
  doi={https://doi.org/10.1103/PhysRevB.75.045404},
  publisher={APS}
}

@article{novoselov2004,
  title={Electric field effect in atomically thin carbon films},
  author={Novoselov, Kostya S. and Geim, Andre K. and Morozov, Sergei V. and Jiang, De-eng and Zhang, Yanshui and Dubonos, Sergey V. and Grigorieva, Irina V. and Firsov, Alexandr A.},
  journal={Science},
  volume={306},
  number={5696},
  pages={666--669},
  year={2004},
  doi={https://doi.org/10.1126/science.1102896},
  publisher={American Association for the Advancement of Science}
}

@article{parra2016,
  title={Stokes--anti-{S}tokes correlation in the inelastic scattering of light by matter and generalization of the bose-einstein population function},
  author={Parra-Murillo, Carlos A. and Santos, Marcelo F. and Monken, Carlos H. and Jorio, Ado},
  journal={Phys. Rev. B},
  volume={93},
  number={12},
  pages={125141},
  year={2016},
  doi={https://doi.org/10.1103/PhysRevB.93.125141},
  publisher={APS}
}

@article{piscanec2004,
  title={Kohn anomalies and electron-phonon interactions in graphite},
  author={Piscanec, S. and Lazzeri, M. and Mauri, Francesco and Ferrari, A. C. and Robertson, J.},
  journal={Phys. Rev. Lett.},
  volume={93},
  number={18},
  pages={185503},
  year={2004},
  doi={https://doi.org/10.1103/PhysRevLett.93.185503},
  publisher={APS}
}

@article{pop2012,
   author = {Pop, Eric and Varshney, Vikas and Roy, Ajit K.},
   title = {Thermal Properties of Graphene: Fundamentals and Applications},
   journal = {MRS Bull.},
   volume = {37},
   number = {12},
   pages = {1273-1281},
   ISSN = {1938-1425},
   DOI = {https://doi.org/10.1557/mrs.2012.203},
   url = {https://doi.org/10.1557/mrs.2012.203},
   year = {2012},
   type = {Journal Article}
}

@article{sullivan2017,
  title={Optical generation and detection of local nonequilibrium phonons in suspended graphene},
  author={Sullivan, Sean and Vallabhaneni, Ajit and Kholmanov, Iskandar and Ruan, Xiulin and Murthy, Jayathi and Shi, Li},
  journal={Nano Lett.},
  volume={17},
  number={3},
  pages={2049--2056},
  year={2017},
  doi = {https://doi.org/10.1021/acs.nanolett.7b00110},
  publisher={ACS Publications}
}

@article{sun2017,
   author = {Sun, Yingying and Yanagisawa, Masahiro and Homma, Takayuki},
   title = {Thermal stability of single-layer graphene subjected to confocal laser heating investigated by using in situ anti-{Stokes} and {Stokes} {R}aman spectroscopy},
   journal = {Electrochemistry},
   volume = {85},
   number = {4},
   pages = {195-198},
   ISSN = {1344-3542},
   DOI = {https://doi.org/10.5796/electrochemistry.85.195},
   year = {2017},
   type = {Journal Article}
}

@article{tristant2016,
  title={Enlightening the ultrahigh electrical conductivities of doped double-wall carbon nanotube fibers by {R}aman spectroscopy and first-principles calculations},
  author={Tristant, Damien and Zubair, Ahmed and Puech, Pascal and Neumayer, Fr{\'e}d{\'e}ric and Moyano, S{\'e}bastien and Headrick, Robert J. and Tsentalovich, Dmitri E. and Young, Colin C. and Gerber, Iann C. and Pasquali, Matteo and Kono, Junichiro and Leotin, Jean},
  journal={Nanoscale},
  volume={8},
  number={47},
  pages={19668--19676},
  year={2016},
  doi={https://doi.org/10.1039/c6nr04647a},
  publisher={Royal Society of Chemistry}
}

@article{weigl2026,
   author = {Weigl, Leonard and Gradl, Johannes and Richter, Peter and Convertino, Domenica and Forti, Stiven and Seyller, Thomas and Coletti, Camilla and Gierz, Isabella},
   title = {Influence of Doping on Non-Equilibrium Carrier Dynamics in Graphene},
   journal = {2D Mater.},
   volume = {13},
   number = {1},
   pages = {015010},
   ISSN = {2053-1583},
   DOI = {https://doi.org/10.1088/2053-1583/ae1bb6},
   url = {https://iopscience.iop.org/article/10.1088/2053-1583/ae1bb6},
   year = {2026},
   type = {Journal Article}
}

@article{Wu2012,
   author = {Wu, Shiwei and Liu, Wei-Tao and Liang, Xiaogan and Schuck, P. James and Wang, Feng and Shen, Y. Ron and Salmeron, Miquel},
   title = {Hot Phonon Dynamics in Graphene},
   journal = {Nano Lett.},
   volume = {12},
   number = {11},
   pages = {5495-5499},
   ISSN = {1530-6984},
   DOI = {https://doi.org/10.1021/nl301997r},
   url = {https://doi.org/10.1021/nl301997r},
   year = {2012},
   type = {Journal Article}
}

@article{yan2007,
  title={Electric field effect tuning of electron-phonon coupling in graphene},
  author={Yan, Jun and Zhang, Yuanbo and Kim, Philip and Pinczuk, Aron},
  journal={Phys. Rev. Lett.},
  volume={98},
  number={16},
  pages={166802},
  year={2007},
  doi={https://doi.org/10.1103/PhysRevLett.98.166802},
  publisher={APS}
}


\newpage




\includepdf[]{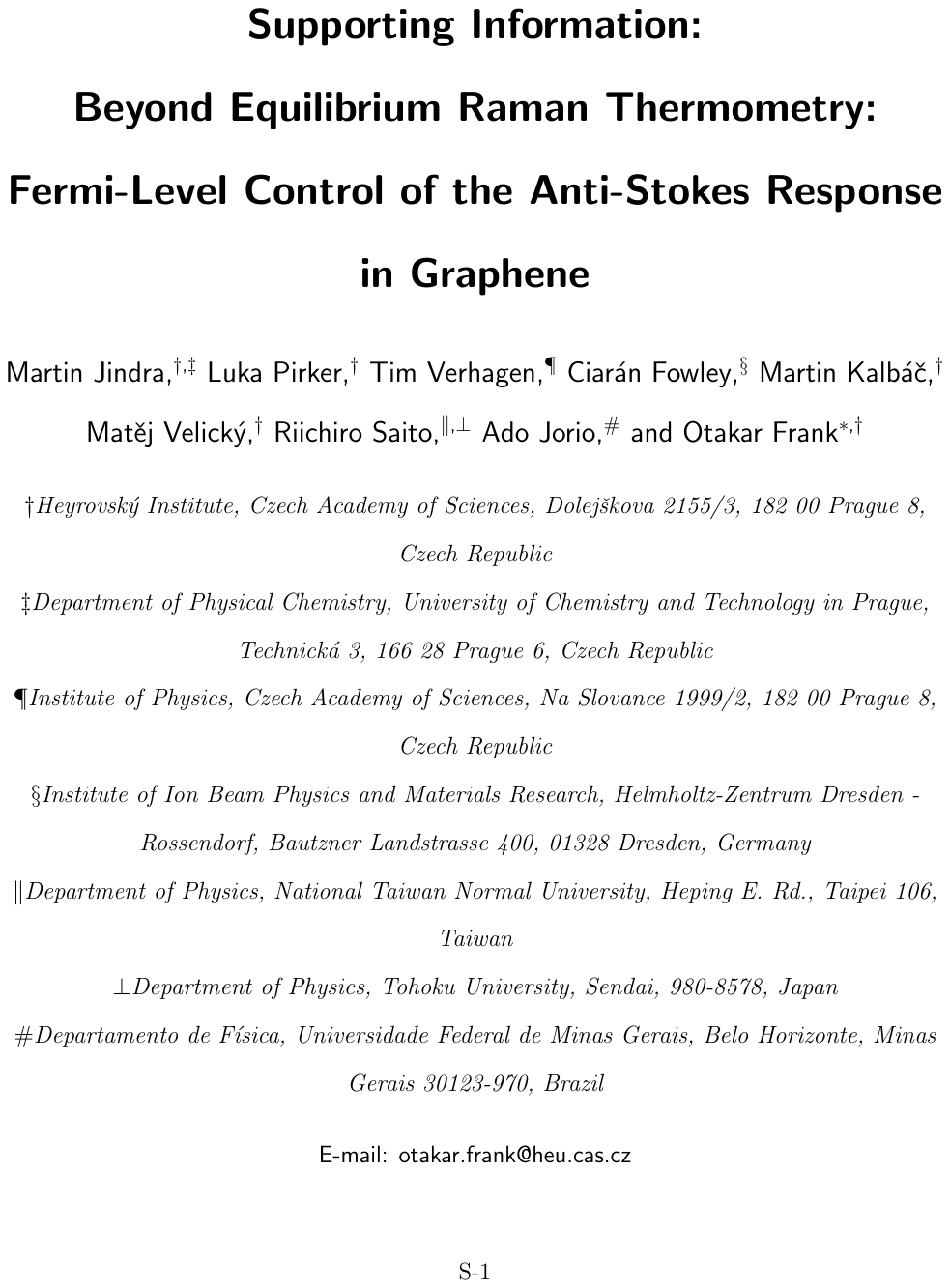}


\setcounter{figure}{0}
\renewcommand{\thefigure}{S\arabic{figure}}

\setcounter{table}{0}
\renewcommand{\thetable}{S\arabic{table}}

\setcounter{section}{0}
\renewcommand{\thesection}{S\arabic{section}}

\setcounter{subsection}{0}
\renewcommand{\thesubsection}{\thesection.\arabic{subsection}}

\setcounter{page}{2}
\renewcommand{\thepage}{S-\arabic{page}}




\newpage



\newpage


\newpage

\section{Sample preparation}
\label{Sample_prep-SI}
\subsection{Mechanical exfoliation}

Monolayer graphene was exfoliated from graphite crystals (NGS Graphit GmbH) using a 1004R-11.0 tape (USI Ultron Systems Inc.). Silicon wafers with a 90 nm SiO\textsubscript{2} layer were used as substrates and cleaned in an ultrasonic bath (Elmasonic P, Elma Schmidbauer GmbH) using acetone and isopropyl alcohol (Microchemicals GmbH) before exfoliation. The wafers were then exposed to oxygen plasma for 60 s (1000 W, Diener electronic GmbH). The tape with bulk graphite was brought into contact with the wafer, macroscopic air bubbles were removed with a cotton swab, and the wafer was heated on a hot plate for 2 min at 100 \textcelsius, before peeling the tape off. Graphene monolayers were identified by optical contrast, the intensity ratio of the Raman G and 2D bands, and the 2D band width.

\subsection{Lithographic patterning for electrolyte gating}\label{basins}

The lithographic resist ECI 3007 (Microchemicals GbmH) was spin-coated onto the wafer with exfoliated graphene at 2000 rpm for 60 s and baked at 90 \textcelsius{} for 60 s. The disc pattern was projected onto the sample with MicroWriter ML3 Pro (Durham Magneto Optics), after which the sample was baked again at 110 \textcelsius{} for 90 s. The resist was developed using AZ 2026 MIF solution (Microchemicals GmbH). The final resist thickness was determined by spectroscopic ellipsometry to be 1 \textmu m.

\subsection{Electron-beam-patterned contact deposition for back-gated sample}\label{ebeam}

Metallic contacts to exfoliated graphene were made via electron beam lithography and lift-off. First, a bilayer resist consisting of Copolymer EL11 (MicroChem) and PMMA A4  (MicroChem) was spun at 4000 rpm and baked at 150 \textcelsius{} for 5 minutes on a hotplate. Contact designs were exposed in a Raith 150 two electron-beam lithography tool at an accelerating voltage of 10 kV. After exposure, samples were developed in a 7:3 solution of isopropanol and deionized water. A Creavac Creamet 600 evaporator with a base pressure of less than $1\times 10^{-7}$ mbar was used for metal film deposition. First, a 5 nm Ti adhesion layer was deposited by electron beam deposition with a growth rate of 1 \AA s$^{-1}$, then, without breaking vacuum, a 75 nm Au contact layer was deposited by thermal evaporation. Both layers were deposited in a vacuum of less than $5\times 10^{-6}$ mbar.

\newpage
\section{Measurements}
\label{Measurement-SI}
\subsection{Raman spectroscopy}

Raman spectra were measured with a WITec Alpha 300R spectrometer (Oxford Instruments) equipped with a filter capable of measuring both the Stokes and anti-Stokes regions of the spectrum using a 633 nm excitation wavelength and a 600 grooves/mm grating. The ultralong-working-distance objective with 100$\times$ magnification (NA 0.7, Mitutoyo) was used for spectroelectrochemical and backgating measurements. The sample mapping and measurements on the chemically gated samples were performed with a 100$\times$ objective (NA 0.9, Zeiss EC Epiplan-Neofluar). The laser power under the objective was adjusted with a neutral-density filter and measured using the PM100D digital optical power meter (ThorLabs, Inc.). All Raman measurements were performed under continuous-wave laser excitation, with a laser excitation wavelength of 633~nm, at 1.5~mW power unless stated otherwise.

To resolve small changes in peak intensities, we performed a spectral intensity calibration. The raw Raman spectra were corrected using a white-light reference spectrum (HL-2000-CAL, Ocean Optics). For each spectral region (Stokes and anti-Stokes), we computed a calibration curve as the ratio of the parametric lamp spectrum to the measured white-light spectrum after background subtraction and multiplied this curve by the background-subtracted raw Raman spectrum to obtain the calibrated spectrum. The Stokes and anti-Stokes regions were acquired with different integration times to detect the weaker anti-Stokes signal while avoiding saturation in the Stokes region (primarily due to the Si signal, which was used for spectral position calibration). The two regions were recorded with a spectral overlap, which was used to determine a scaling factor so that peak intensities could be compared despite different integration times.

\subsection{Electrolyte gating}

Electrolyte gating was performed using a custom-built microdroplet spectroelectrochemical setup, enabling simultaneous electrostatic control of the graphene Fermi level and Raman spectrum monitoring. A small electrolyte droplet of 3.8 M LiClO\textsubscript{4} or 6 M LiCl aqueous electrolyte was expelled into the lithographically manufactured reservoirs on monolayer graphene (described in section \ref{basins}) with a glass capillary, pulled from a borosilicate glass tube using a P-1000 micropipette puller (both Sutter Instrument), whose movement was operated by a 3-axis motorized manipulator (Siskiyou Corp.) The resulting electrolyte reservoirs ranged from 10 to 20 \textmu m in diameter and had approximate heights of 1 \textmu m. Graphene flakes were electrically contacted using a silver conductive paint (Electrolube).

We employed a three-electrode configuration, with the electrolyte/graphene interface as the working electrode, and the reference (Ag/AgCl wire) and counter (Pt wire) electrodes (both Goodfellow Inc.) placed inside the glass capillary. The potential applied to graphene with respect to the reference electrode was controlled with CompactStat.h10800 potentiostat (Ivium Technologies B.V.), allowing for reversible tuning of the Fermi level. The electrochemical window was restricted to avoid undesirable Faradaic processes, ensuring that the observed spectral changes arise predominantly from electrostatic doping rather than chemical changes in graphene or electrolyte.

\subsection{Dielectric back-gating}

Dielectric gating was performed using a back-gated device geometry. A voltage was applied (Keithley 2612B) between a highly doped silicon back-gate and monolayer graphene contacted by electron beam lithography (section \ref{ebeam}), which were separated by a 90~nm thick SiO\textsubscript{2}. 

\subsection{Chemical doping}

Chemical doping was carried out following the protocol reported by Choi \textit{et al.},$^1$ in which a 10~mM aqueous solution of AuCl$_3$ (Sigma-Aldrich, 99.99$\%$ purity) was spin-coated onto exfoliated graphene, followed by thermal annealing at 80 \textcelsius. Laser power used for the measurement of these samples was 0.5~mW.


\newpage
\section{Data analysis}

\subsection{Peak fitting}
\label{SI_peakfitting}
Raman spectra were analyzed using a multi-component fitting procedure implemented in a customized Python-based framework. The spectral regions corresponding to the Raman Si, G, and 2D bands were fitted using non-linear least-squares optimization with a linear background approximation. The Si and G bands were modeled using a Breit-Wigner-Fano (BWF) line shape to account for the interaction between the discrete optical phonon and the electronic continuum,$^2$ which results in an asymmetric peak shape. The BWF function is given by:

\begin{equation}
A(\omega) = A_0 \frac{(1 + s/q_{\textrm{BWF}})^2}{1 + s^2},
\end{equation}

where

\begin{equation}
s = \frac{\omega - \omega_0}{\Gamma}.
\end{equation}

Here, $\omega$ is the Raman shift, $\omega_0$ is the peak position, $\Gamma$ is the linewidth parameter, $1/q_{\mathrm{BWF}}$ is the BWF asymmetry factor, and $A_0$ is the peak amplitude. The peak intensity was obtained as an integrated area of the optimized function.\\




In the case of the G band, the uncertainty of the spectral intensity was included in the fitting procedure, assuming photon-counting (Poisson) statistics for the raw detector counts. The standard deviations of the measured and background intensities at each spectral point (i) were therefore estimated as

\begin{equation}
\sigma_{\mathrm{meas},i}=\sqrt{N_{\mathrm{meas},i}},
\qquad
\sigma_{\mathrm{bg},i}=\sqrt{N_{\mathrm{bg},i}},
\end{equation}

where $N_{\mathrm{meas},i}$ and $N_{\mathrm{bg},i}$ denote the measured sample and background counts, respectively. The corresponding uncertainties were propagated through the background subtraction and subsequent intensity calibration, including the uncertainties associated with the experimentally determined calibration function and Stokes/anti-Stokes scaling coefficient. The resulting point-wise uncertainties of the corrected spectral intensity were used as absolute standard deviations in the weighted non-linear least-squares optimization. Consequently, the fit minimized the uncertainty-weighted residuals, such that spectral points with larger uncertainties contributed proportionally less to the fitted model.

The 2D band was fitted using a Lorentzian line shape:

\begin{equation}
A(\omega) = \frac{A_0}{1 + \left( \frac{\omega - \omega_0}{\Gamma} \right)^2},
\end{equation}

where $\omega$ is the Raman shift, $\omega_0$ is the peak position, $\Gamma$ is the linewidth parameter, and $A_0$ is the peak amplitude. The peak intensity was obtained as an integrated area of the optimized function. Signal uncertainty was not included for the 2D band.\\

\subsection{Correlation analysis}\label{corr}

To decouple the respective contributions of strain and charge doping in graphene, a correlation analysis of the 2D- and G-band Raman peak positions was carried out. This approach exploits the distinct and well-characterized sensitivities of these vibration modes to mechanical deformation (strain) and charge carrier concentration (doping), as quantitatively described in detail by Lee \textit{et al.}.$^3$ The analysis was carried out in the $(\omega_{\mathrm{2D}}, \omega_{\mathrm{G}})$ space with approximately linear correlations with slopes of approximately 2.2 (strain) and 0.7 (charge doping), which allowed us to decompose the experimental spectral shifts into strain and doping components. The origin of the coordinate system is defined using the reference position of undoped, unstrained graphene and recalculated for the 633~nm excitation wavelength.$^3$

\newpage
\section{Supplementary data}

\subsection{Characterization of the lithographically patterned sample}

\begin{figure}[h!]
    \centering
    \includegraphics[width=1\linewidth]{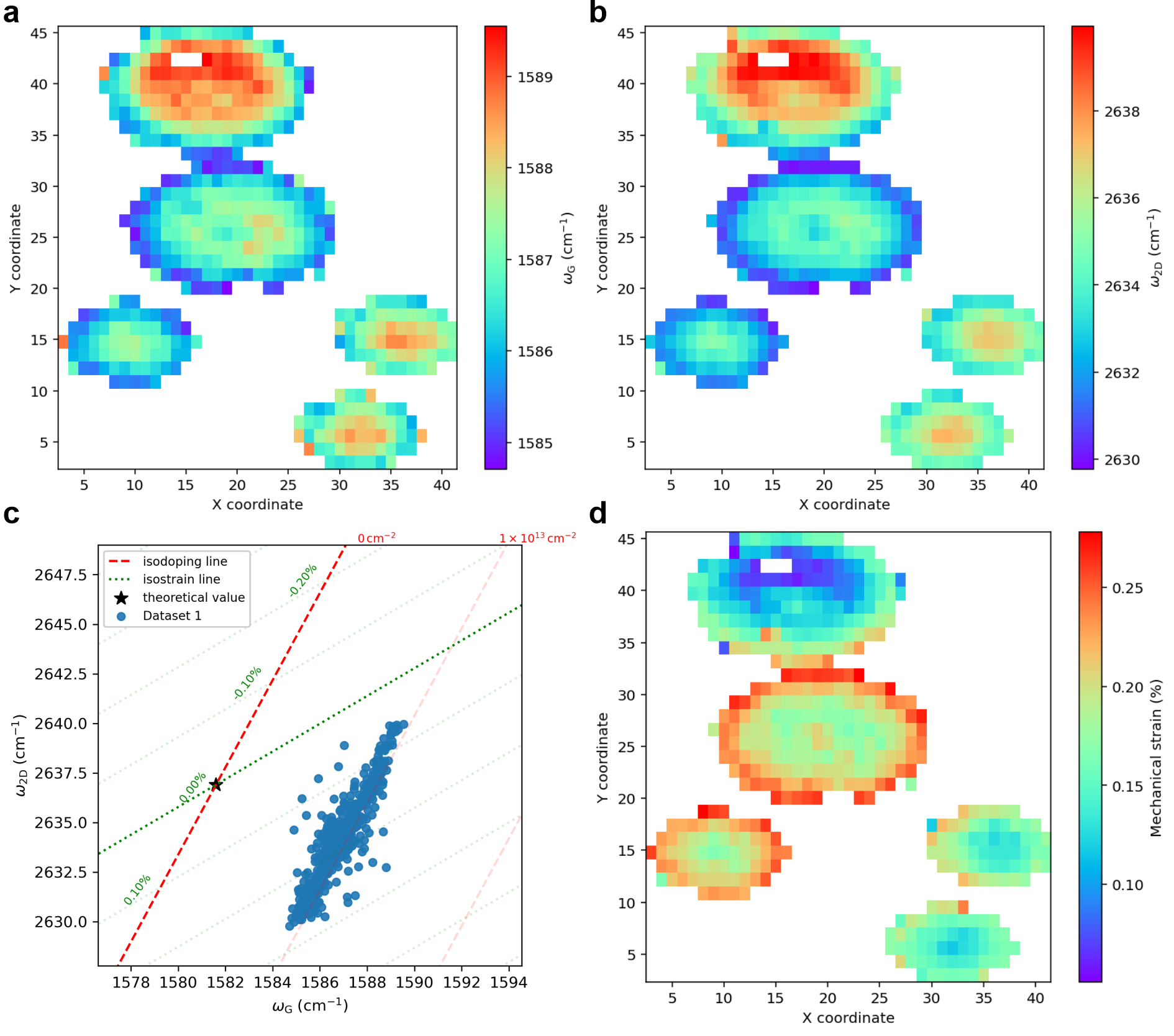}
    \caption{\textbf{Raman G and 2D peak maps and correlation analysis.} \textbf{a, b}, Spatial distribution of the G and 2D band position, respectively, \textbf{c,} Correlation of the 2D and G band position, with the isostrain and isodoping lines shown, \textbf{d,} Distribution of mechanical strain in the sample extracted from the correlation (distribution of charge carrier concentration is presented in the main text).}
    \label{map_corr}
\end{figure}

\newpage

\subsection{2D--G mode correlation analysis with variable gate voltage}

\begin{figure}[h!]
    \centering
    \includegraphics[width=0.75\linewidth]{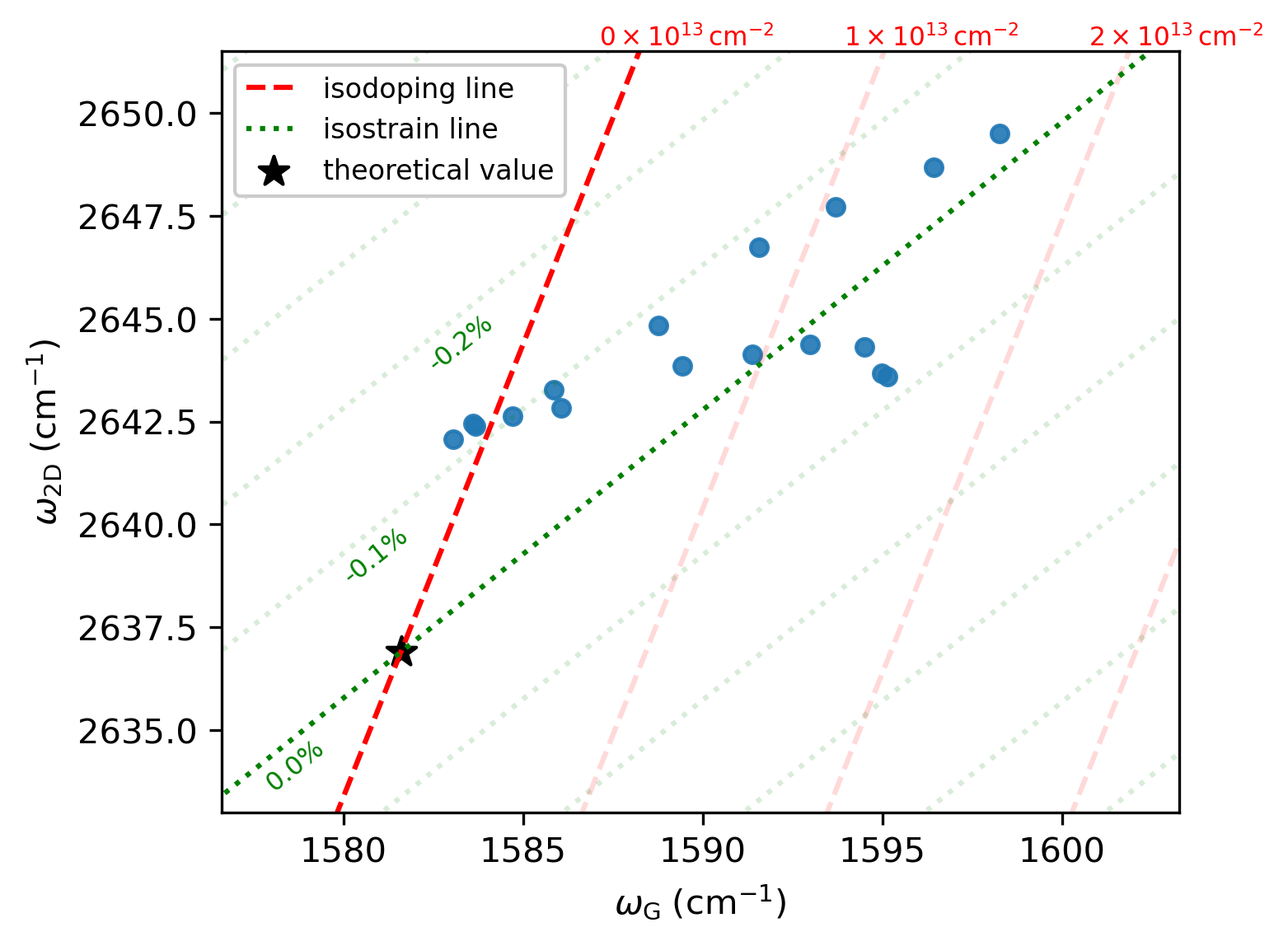}
    \caption{\textbf{2D–G peak correlation under electrolyte gating.} Correlation analysis of the electrolyte gating presented in the main text in Fig. 2 showing the evolution along the isostrain line caused by doping, and the difference between electron and hole doping.}
    \label{corr_single}
\end{figure}

\newpage

\subsection{Evolution of the G band parameters with gate voltage}

\vspace{-9pt}

\begin{figure}[h!]
    \centering
    \includegraphics[width=0.75\linewidth]{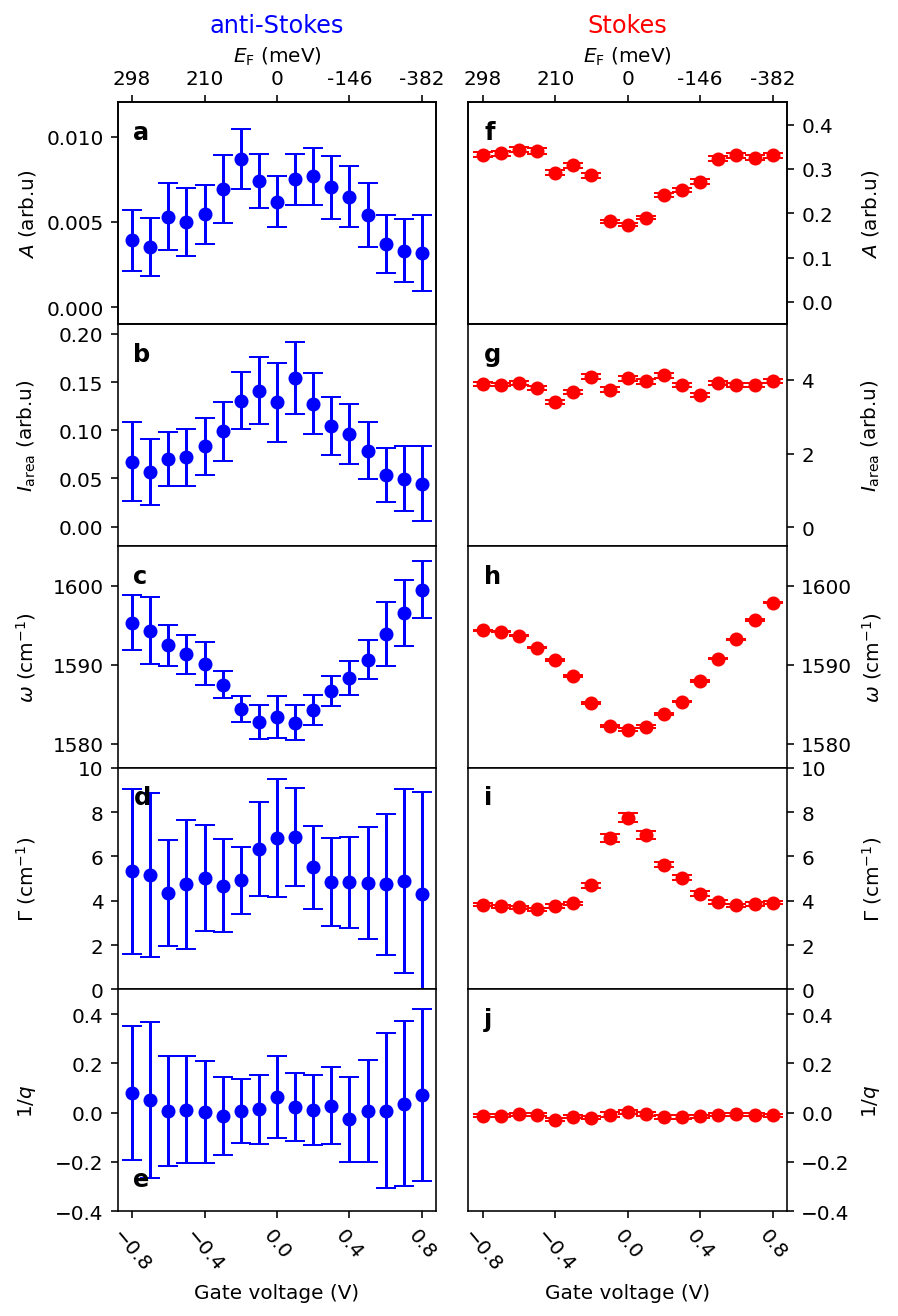}
    \caption{\textbf{BWF lineshape analysis of the Raman G band.} G band parameters obtained by Fano lineshape fitting. \textbf{a--e,} Amplitude, areal intensity (calculated from other peak parameters; for a comparison with the integrated areas plotted in Fig.~2b, main text), position, width, and 1/\textit{q} in the anti-Stokes region, respectively. \textbf{f--j,} Same parameters as a--e for the Stokes region.}
    \label{all_G}
\end{figure}

\newpage

\subsection{Comparison of various gating approaches}

\vspace{-9pt}

\subsubsection{Dielectric back-gating}

\vspace{-9pt}

\begin{figure}[h!]
    \centering
    \includegraphics[width=0.9\linewidth]{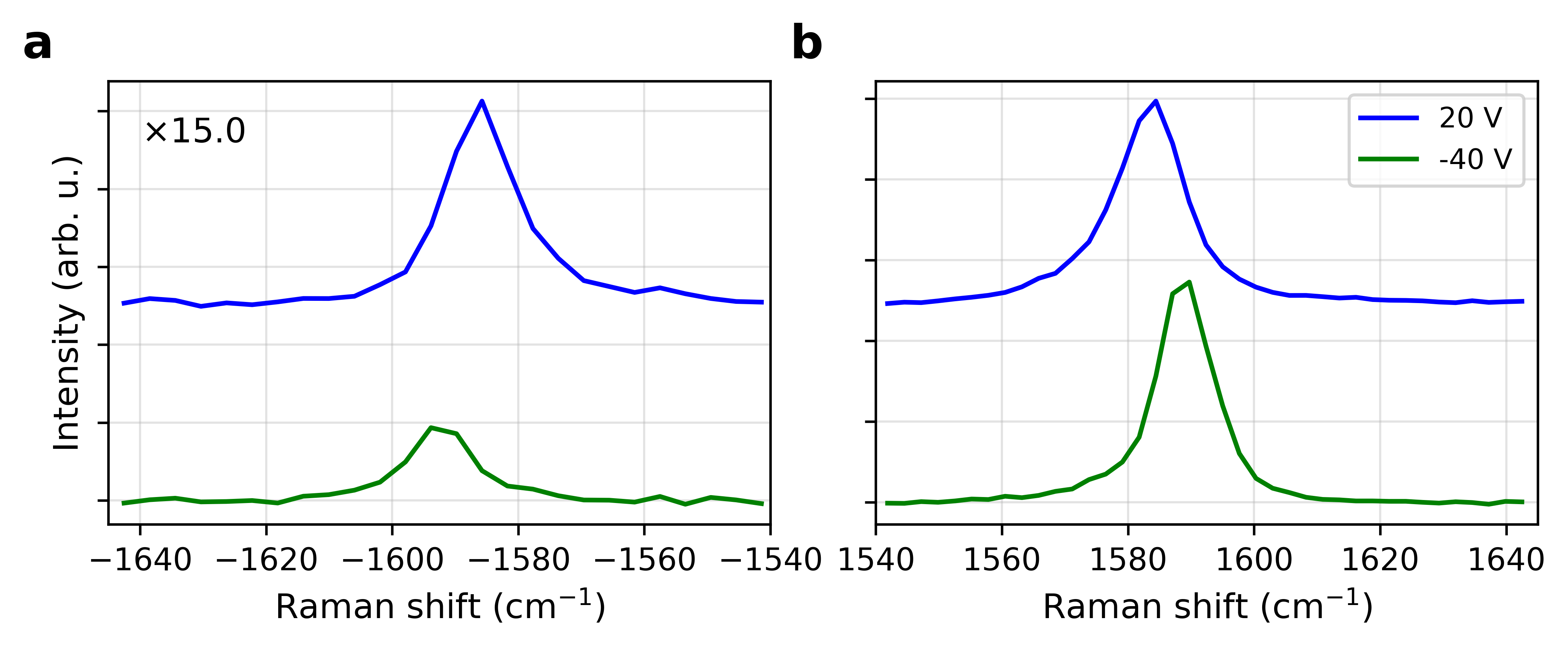}
    \caption{\textbf{Graphene dielectric back-gating.} Spectra of graphene with applied gate voltage of 20 (close to Dirac point) and -40~V (maximum applied voltage) in anti-Stokes (\textbf{a}) and Stokes (\textbf{b}) regions. The Stokes spectra show a significantly upshifted G band as the gate voltage shifts the Fermi level away from the Dirac point, with no major change in peak intensity. The anti-Stokes spectra show a mirrored shift but a significant reduction in intensity at the same applied voltage.}
    \label{fig:backgate}
\end{figure}

\vspace{-9pt}

\subsubsection{Chemical gating}

\vspace{-9pt}

\begin{figure}[h!]
    \centering
    \includegraphics[width=0.9\linewidth]{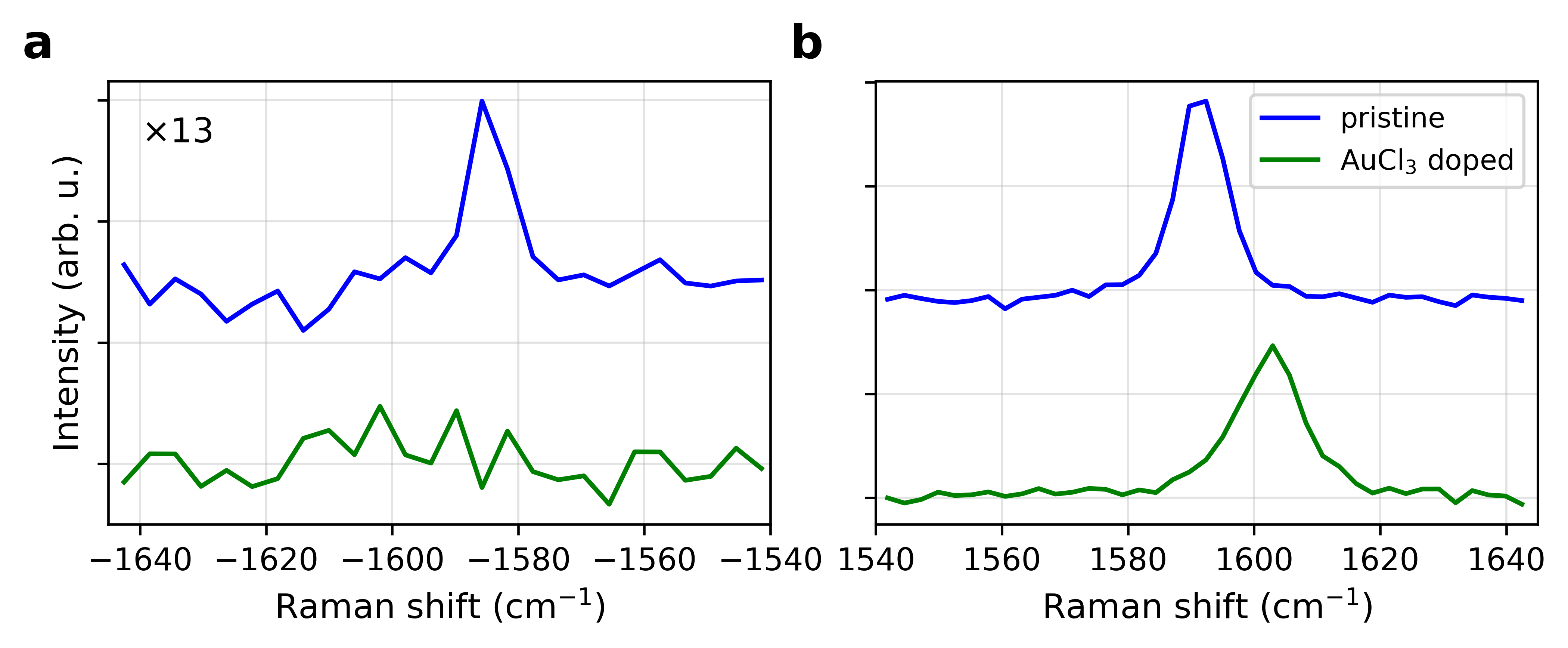}
    \caption{\textbf{Graphene chemical doping.} Spectra of pristine and chemically doped graphene G band in anti-Stokes (\textbf{a}) and Stokes (\textbf{b}) region. Stokes spectra show a significantly upshifted G band after $\mathrm{AuCl_3}$ exposure, without a major change in peak intensity. The anti-Stokes spectra show a lowering of the anti-Stokes signal upon chemical doping.}
    \label{fig:chemgate}
\end{figure}

\newpage

\subsubsection{anti-Stokes/Stokes ratio during gating}

\begin{figure}[h!]
    \centering
    \includegraphics[width=0.7\linewidth]{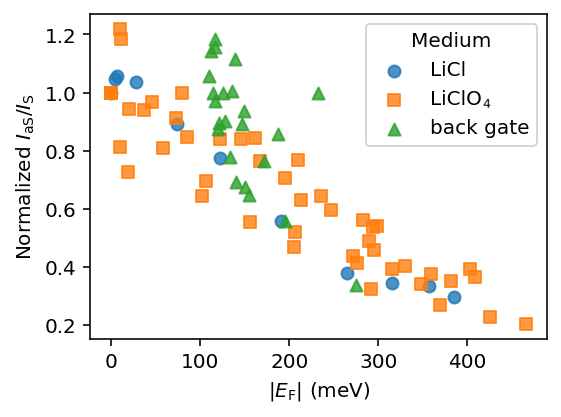}
    \caption{Anti-Stokes/Stokes areal intensity ratio of datasets measured with the LiCl and $\mathrm{LiClO_{4}}$ electrolyte gates, and the dielectric back gate on samples with electron-beam deposited electrodes. For an easier comparison of the overall trends, the $I_{\mathrm{aS}}/I_{\mathrm{S}}$ is normalized to the calibrated $I_{\mathrm{aS}}/I_{\mathrm{S}}$ measured at the Dirac point separately for each dataset.}
    \label{gating_all}
\end{figure}

\newpage

\subsection{Doping dependence of the G-band width and anti-Stokes/Stokes ratio: Electrolyte gating datasets}

\begin{figure}[h!]
    \centering
    \includegraphics[width=1\linewidth]{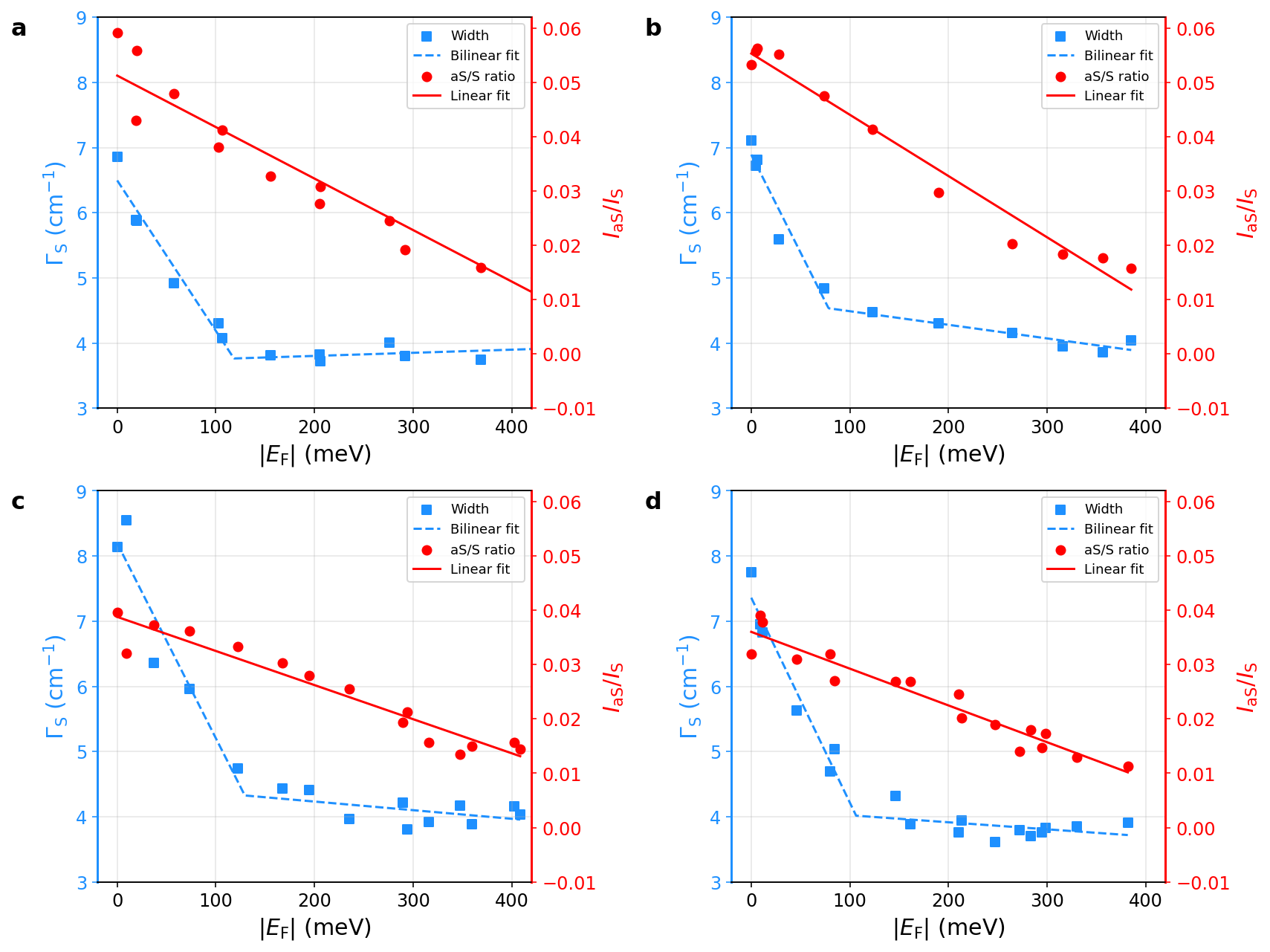}
    \caption{\textbf{G-band width and $I_{\mathrm{aS}}/I_{\mathrm{S}}$ evolution.} Datasets collected at 4 different measurements showing the $|E_{\mathrm{F}}|$-dependence of G-band width with a bilinear fit (blue, left axis) and $I_{\mathrm{aS}}/I_{\mathrm{S}}$ with a linear fit (red, right axis). \textbf{a,} Data measured using 3.8~M aq. solution of $\mathrm{LiClO_4}$ and 633~nm laser at 2~mW, \textbf{b,} data measured using 6~M aq. solution of $\mathrm{LiCl}$ and 633~nm laser at 2~mW , \textbf{c,} data measured using 3.8~M aq. solution of $\mathrm{LiClO_4}$ and 633~nm laser at 2~mW, \textbf{d,} data measured using 3.8~M aq. solution of $\mathrm{LiClO_4}$ and 633~nm laser at 1.5~mW.}
    \label{bilinear_fitting_sec}
\end{figure}

\newpage

\includepdf[]{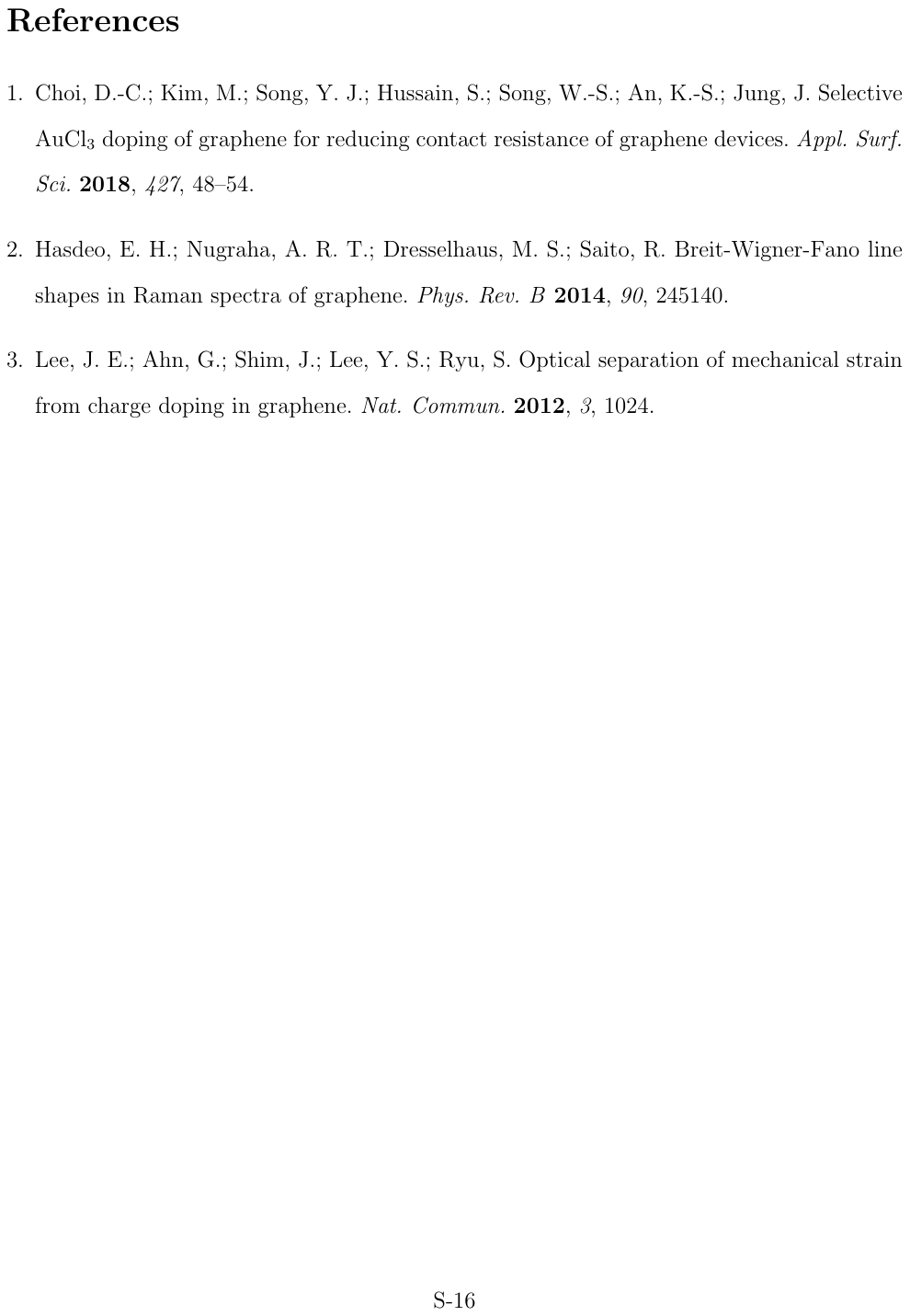}


\end{document}